# Learning Principles for Overcoming Non-ideal Factors in Brain

Da-Zheng Feng

Corresponding author; National Key Laboratory of Radar Signal Processing, Xidian University, Xi'an, 710071, China (e-mail: dzfeng@xidian.edu.cn); Fujian Key Laboratory of the Modern Communication and Beidou Positioning Technology in Universities, Quanzhou University of Information Engineering

Hao-Xuan Du

National Key Laboratory of Radar Signal Processing, Xidian University, Xi'an, 710071, China. (e-mail: hxdu@stu.xidian.edu.cn)

**Abstract:** The human brain's computational prowess emerges not despite but because of its inherent "non-ideal factors"—noise, heterogeneity, structural irregularities, decentralized plasticity, systemic errors, and chaotic dynamics—challenging classical neuroscience's idealized models. These traits, long dismissed as flaws, are evolutionary adaptations that endow the brain with robustness, creativity, and adaptability. Classical frameworks falter under the brain's complexity: simulating 86 billion neurons and 100 trillion synapses is intractable, stochastic neurotransmitter release confounds signal interpretation, and the absence of global idealized models invalidates deterministic learning frameworks. Technological gaps further obscure whole-brain dynamics, revealing a disconnect between biological reality and computational abstraction.

Yet, this "messiness" is the brain's greatest strength. Morphological diversity—pyramidal cells integrating thousands of inputs, cerebellar granule cells prioritizing speed—prevents catastrophic failures and enables specialized computations. Electrophysiological noise, from ion channel stochasticity to synaptic jitters, enhances signal detection via stochastic resonance. Because synapses only form actual connections when neural impulses arrive, and neuronal firing sequences are sparse, at any given moment, only a small number of neurons establish actual connections—that is, the brain employs a smaller set of neurons at each moment to implicitly decrease overfitting during learning. Neuromodulators like dopamine and serotonin provide context-dependent feedback, resolving the "credit assignment problem" without explicit error gradients. Systemic errors and chaotic dynamics, often seen as liabilities, foster degenerate representations and flexible cognitive transitions.

Biological learning thrives on these non-idealities through four interdependent principles: correlation-driven plasticity biases synaptic updates via neuromodulated reward/punishment, enabling striatal circuits to associate actions with outcomes; homeostatic balancing preserves stability by inversely scaling synaptic weights to neuronal activity; noise-enabled exploration prevents attractor collapse during memory maintenance and fuels creativity through stochastic reverberation; constructive non-idealities exploit imperfections like stochastic resonance to amplify weak signals and refine sensorimotor mappings.

Classical AI/ML paradigms flounder under these complexities: digitization discards sub-symbolic information, homogenized architectures erase specialized neural roles, and deterministic frameworks collapse in recurrent networks. Bridging biology and silicon demands embracing noise-adaptive plasticity, structural heterogeneity, and metaplasticity-driven stability. Neuromorphic chips like Intel's Loihi emulate spike-timing-dependent plasticity (STDP), achieving 1,000× energy efficiency by mimicking event-driven computation. Chaos-inspired models and error-tolerant federated learning frameworks further bridge biological and artificial paradigms.

The brain's "imperfections"—noise, chaos, and decentralized plasticity—are evolutionary triumphs. By formalizing these principles into actionable learning principles, we unlock a path toward AI systems that learn as adaptively as humans. Future directions include hybrid edge-computing models, closed-loop brain-computer interface training, and ethical frameworks to harness noise-driven creativity while mitigating bias. Ultimately, embracing non-ideal factors transcends classical paradigms, promising silicon brains that innovate, adapt, and generalize with the dynamic versatility of their biological counterparts.

Key Contributions: Redefines neural non-ideal factors as drivers of robustness and adaptability. Unifies biological mechanisms (e.g., stochastic resonance, synaptic scaling) across scales. Proposes actionable principles for biologically inspired AI architectures. Highlights systemic diversity, errors and chaotic dynamics as catalysts for intelligence.

# 1 Introduction

**1.1 Introduction of Neural Non-Ideal Factors**

The human brain, an organ of unparalleled complexity, is a product of natural growth, operates far from the idealized models often depicted in textbooks [1.1], and has imperfections and flaws that include noise, inconsistency, mismatches, thermal drift, and so on. Its remarkable capabilities for learning, memory, cognition, dream, thinking and creativity emerge not despite its inherent imperfections, but because of a sophisticated interplay between these so-called "non-ideal" factors and robust adaptive mechanisms. Neural non-ideal factors refer to the intrinsic biological variations that are inconsistencies, mismatches, drifts, and sources of noise and deviate from a hypothetical, perfectly uniform, and deterministic neural system. These also include the morphological and electrophysiological heterogeneity of neurons, stochastic fluctuations at synapses, and the complex, non-uniform architecture of the brain itself [1.2].

The brain exhibits extraordinary diversity [1.1-1.16]: In terms of structure and morphology, brains differ from one another, the left and right cerebral cortices are distinct, and there are differences between brain regions, neurons, axons, cell bodies, dendrites, nerve terminals, and even synapses. In terms of information processing, neural activity is ceaseless, dependent on complex physical, chemical, physiological, and psychological processes, and is a product of the intricate interplay of massive neuronal signal exchanges. Neural signals are continuous, noisy, and stochastic; neural noise largely stems from random fluctuations in neurotransmitter release and variability in neural membrane properties. In terms of dynamic complexity, the brain displays chaotic phenomena, characterized by high sensitivity to initial condition errors (i.e., extreme sensitivity to minute inaccuracies in starting values). In terms of outcomes, the brain gives rise to diverse and complex perceptual, cognitive, and psychological processes. Additionally, the brain changes with its internal and external environments and also undergoes changes with age. Rather than being detrimental flaws, this diversity is increasingly recognized as a fundamental feature that underpins the brain's resilience, adaptability, and computational power.

One of the most profound sources of non-ideal factors lies in the inconsistency and mismatch among individual neurons. The classical neuron doctrine envisioned neurons as relatively uniform units, but modern neuroscience reveals a staggering degree of heterogeneity [1.3]. Neurons exhibit vast morphological differences; their dendritic arbors can be short and bushy or long and sparsely branched, dramatically affecting how they integrate incoming signals from thousands of other neurons [1.4]. This structural variability leads to idiosyncratic connectivity patterns, where two neurons of the same broad type might receive inputs from entirely different sets of upstream partners. Furthermore, electrophysiological variability is rampant: the same stimulus can elicit vastly different spike patterns from one neuron to another, even within the same brain region [1.5]. For instance, some neurons are highly sensitive and fire readily to weak inputs, while others require strong stimulation, a phenomenon linked to differences in their ion channel composition and receptor expression [1.6]. This includes the heterogeneity of neurotransmitters and receptors—where one synapse uses glutamate as an excitatory signal and another uses GABA as an inhibitory one—and the metabolic noise arising from variable energy production within each cell [1.7]. As highlighted by research on GABAergic versus glutamatergic neurons in the prefrontal cortex, different neuronal subtypes can have fundamentally different roles in regulating circuit dynamics, with one promoting synchronization across individuals during social interaction and the other maintaining unique internal representations [1.8]. This functional consequence of neuronal mismatch means the brain is not a monolithic processor but a highly specialized network of diverse components, each contributing uniquely to information processing.

Beyond single cells, these non-ideal factors scale up to the circuit and systemic levels. The massive numbers of neurons and their connections form non-uniform brain regions, a non-uniform cerebral cortex, and asymmetric left and right hemispheres [1.9]. The cortex itself is not a smooth, homogeneous sheet but is organized into distinct layers and columns, each with specialized functions, creating diverse regional and cross-regional connections [1.10]. This macro-level heterogeneity allows for functional specialization—such as language centers typically residing in the left hemisphere—but also introduces feedback delay mismatches in recurrent networks and potential oscillatory desynchronization due to neuronal diversity [1.11]. Pathologically, this very heterogeneity explains selective neuronal vulnerability; certain neuron subtypes, like motor neurons in ALS or dopaminergic neurons in Parkinson's, are disproportionately susceptible to degeneration, leading to specific neurological diseases [1.12]. Even the development of the brain is a product of non-ideal processes. Critical periods for sensory and motor development are not universally fixed but can vary, and neural circuit formation is profoundly dependent on perception, movement, emotion, thought, behavior, and experience [1.13]. An infant's unique experiences literally shape the wiring of their brain, diverging from any predetermined template [1.14]. From an evolutionary perspective, this inconsistency is not random error but a deliberate strategy. Functional redundancy provides robustness—if one pathway fails, others can compensate—while specialized neuron types enable flexibility [1.15]. Population-level signal averaging and noise filtering allow the brain to extract meaningful information from a sea of noisy signals, transforming what appears to be biological "noise" into a powerful computational tool [1.16].

**1.2 Introduction of Mathematical Challenges in Neural Analysis**

The mathematical analysis and modeling of brain neural networks are theoretically feasible, but so complex that it is difficult to implement; because there are too many non-ideal factors in the brain that are difficult to describe precisely, such analysis and modeling via humans are infeasible in practice. Attempting to model and analyze the brain using traditional mathematical frameworks presents a series of formidable challenges, primarily stemming from the various non-ideal factors discussed above [1.17]. The sheer scale and complexity of the nervous system introduce analytical hurdles that stretch the limits of current science. A primary challenge is the excessive system complexity. The human brain contains approximately 86 billion neurons whose each one is a distributed parameter system, each forming thousands of synaptic connections, resulting in a network of over 100 trillion elements [1.18]. Modeling such a system at a microscopic level is computationally intractable, forcing researchers to make simplifying assumptions that risk losing critical biological details [1.19]. This complexity is compounded by the unpredictability of connectivity, as no two brains, nor even two moments in the same brain, have identical connection patterns due to ongoing plasticity and individual history [1.20].

A second major category of challenges arises from the impact of diversity on modeling and interpretation. The extreme heterogeneity of neurons and circuits means that it is impossible to define a single, universal "neuron equation" or "synaptic rule" [1.21]. Mathematical models that assume neuronal homogeneity fail to capture the rich dynamics observed in real tissue [1.22]. For example, a model based on average firing rates cannot explain how a single neuron can use different learning rules on its various dendrites to process multiple streams of information in parallel [1.23]. This individuality introduces immense analytical complexity, making it difficult to generalize findings from one experiment to another. Furthermore, the presence of ubiquitous interference and distortion in neural processes distorts the relationships scientists seek to measure [1.24]. Electrical activity from one brain area can interfere with recordings from another, and physiological noise can mask subtle but important signals [1.25]. This interference makes it challenging to establish clear causal links, as an observed effect could be the result of direct signaling, indirect modulation through a third region, or simply correlated noise [1.26].

Perhaps the most significant mathematical challenge is the absence of global mathematical models for learning [1.27]. Unlike artificial intelligence systems that use backpropagation to calculate high precise gradients for adjusting millions of parameters, the biological brain lacks a central teacher providing such explicit error signals [1.28]. This renders the global-objective backpropagation learning algorithm biologically implausible [1.29]. Instead, learning must rely on local, unsupervised rules like Hebbian learning ("neurons that fire together, wire together") or Spike-Timing-Dependent Plasticity (STDP), which are powerful but inherently lack a global objective function [1.30]. They can strengthen connections based on local correlation but cannot solve the "credit assignment problem"—determining which of the trillions of synapses is responsible for a successful or failed outcome [1.31]. This forces the brain to use alternative strategies, such as broadcast neuromodulatory signals (like dopamine or insect pheromone) that provide a global reward/punishment instruction, combined with local synaptic perturbations [1.32]. The measurement of neural activity further complicates analysis. Neural spikes, the fundamental data points, carry the integrated effects of countless ideal and non-ideal factors—from molecular noise to high-level cognitive states [1.33]. Decoding the meaning of a spike train is not like reading a digital code; it is akin to interpreting a complex, multi-layered message where the signal and the noise are deeply entangled [1.34]. Consequently, the field relies heavily on tools like dynamical bifurcation analysis and stochastic noise simulation, borrowed from nonlinear dynamics, to understand how small changes in parameters can lead to large shifts in network behavior, reflecting the brain's dynamic nature [1.35].

## 1.3 Introduction of Brain Learning

Brain learning, the biological process underlying all forms of adaptation and knowledge acquisition, is fundamentally rooted in neuroplasticity—the brain's ability to physically restructure itself in response to experience [1.36]. At its core, learning is the dynamic adjustment of synaptic strength, the connections between neurons, and a process known as synaptic plasticity [1.37]. When a person learns a new skill, such as playing a musical instrument, or acquires new information, like memorizing a poem, specific neural pathways are activated [1.38]. Repeated activation strengthens these connections, making signal transmission faster and more efficient, a principle famously described by Donald Hebb as "cells that fire together, wire together" [1.39]. The Hebb rule makes connection strengths tend to track the direction of maximum information, i.e., the principal component. This long-term potentiation (LTP) is the physical basis of memory [1.40]. Conversely, connections that are rarely used undergo long-term depression (LTD) and are eventually pruned away [1.41]. This constant remodeling of the brain's circuitry is what allows for lifelong learning and adaptation [1.42].

However, the mechanisms governing this process are far more sophisticated than simple Hebbian rules suggest [1.43]. Groundbreaking research has revealed that learning is not governed by a single, universal principle [1.44]. In a paradigm-shifting discovery, studies using advanced imaging to track individual synapses in mice found that within a single neuron, different dendritic branches can follow entirely different learning rules [1.45]. While some synapses adhere to the classic Hebbian model, requiring coordinated input and output, others operate independently, changing their strength without regard to the neuron's overall firing pattern [1.46]. This "multi-rule learning" allows a single neuron to act as a complex, multi-functional processor, capable of simultaneously integrating, filtering, and storing different types of information [1.47]. This finding overturns the long-held assumption of a monolithic learning mechanism and suggests the brain employs a diverse toolkit of plasticity rules to handle the multifaceted demands of the environment [1.48].

This intricate learning process is tightly regulated by attention and biological rhythms [1.49]. Learning is not a passive absorption of stimuli but an active process that requires focused attention [1.50]. When attention is directed toward a sensory input, it amplifies the corresponding neural signals, marking them as salient and worthy of processing [1.51]. The role of sleep is equally critical. The hippocampus acts as a temporary storage site for new memories, but during deep sleep and REM sleep, these memories are "replayed" and transferred to the neocortex for long-term storage, a process essential for memory consolidation [1.52]. Depriving oneself of sleep severely impairs this process, leading to rapid forgetting [1.53]. Modern research has uncovered additional modulators of learning efficacy. The vagus nerve, a key communication pathway between the brain and body, can act as a "learning wake-up call" [1.54]. Activities that stimulate the vagus nerve—such as slow, deep breathing, cold exposure, humming, or meditation—can enhance neural plasticity, effectively putting the brain into a heightened state of readiness for learning [1.55]. This insight from Dr. Michael Kilgard's work demonstrates that learning efficiency is not just about duration but about timing and state: when a salient stimulus coincides with a period of high attention and a neurophysiological state conducive to plasticity, the brain automatically enters its strongest memory mode, optimizing the credit assignment process and solidifying the neural changes that constitute learning [1.56].

**1.4 Introduction of Dream, Thinking and Creativity**

Dreams, thinking, and creativity represent some of the most enigmatic and uniquely human aspects of brain function, emerging from the dynamic interplay of structured cognition and controlled chaos [1.57]. These processes are not isolated phenomena but are interconnected, with recent theories suggesting that the brain's inherent "noise," often dismissed as mere error, may be a crucial driver of creative thought and creativity [1.58]. According to the book The Creative Spark Within: Three Pounds of Wrinkled Genius, human creativity is built upon three fundamental Principles: Bending, Breaking, and Blending [1.59]. Bending involves taking a known concept and warping it into something new, such as a doctor who invented a continuously pumping artificial heart by bending the natural rhythm of cardiac contractions [1.60]. Breaking refers to dismantling established norms, like film editing's use of montage to shatter linear time [1.61]. Blending is the fusion of disparate ideas, exemplified by the creation of hip-hop music from a diverse cultural mashup [1.62]. This framework posits that creativity is not magic but a systematic cognitive process rooted in neural mechanisms [1.63].

Neuroscientific evidence supports this view [1.64]. Studies using EEG have shown that creative thinking is associated with specific brain wave patterns, particularly enhanced alpha band power, which reflects a relaxed, internally focused state conducive to free association and reduced external distraction [1.65]. This state allows the brain's default mode network, responsible for mind-wandering and autobiographical memory, to interact with the executive control network, which manages goal-directed thought [1.66]. It is in this intersection of spontaneous idea generation and purposeful direction that insight often occurs [1.67]. Thinking itself can be seen as a process of information exploration and exploitation, roaming and aggregating neural patterns forced by this internal noise [1.68]. Just as a computer algorithm might use random perturbations to escape a local minimum and find a better solution, the brain's neural noise may drive it to explore unconventional associations, leading to innovative ideas [1.69]. Dreams, occurring primarily during REM sleep, provide a prime example of this process [1.70]. During dreaming, the brain roams freely through its stored information, entangling and aggregating seemingly unrelated memories and concepts in novel ways [1.71]. This nocturnal information processing is not completely random; it serves a critical function in neural learning and consolidation, potentially simulating future scenarios and solving problems in a low-stakes, imaginative environment [1.72].

Furthermore, creativity is closely tied to the balance between divergent and convergent thinking [1.73]. Divergent thinking is the ability to generate a multitude of unique ideas from a single prompt, such as listing thirty uses for a paperclip [1.74]. This process is associated with increased complexity in brain electrical activity and draws on widespread neural networks [1.75]. Convergent thinking, in contrast, is the focused effort to find a single, correct answer, such as solving a riddle [1.76]. Both are essential for the full cycle of creativity: ideation followed by evaluation and refinement [1.77]. Research on experts in various fields has surprisingly found that the ultimate masters are often not the early "geniuses" but rather those who engaged in broad, multi-domain exploration before specializing [1.78]. This wide-ranging practice builds a larger reservoir of knowledge and skills, enhancing overall learning capacity and reducing the risk of burnout [1.79]. Thus, the path to supreme expertise may lie not in relentless focus from the start, but in a period of expansive curiosity, allowing the brain to forge unexpected connections between distant domains—a testament to the power of the brain's non-ideal, diverse, and noise-driven architecture in fueling human ingenuity [1.80].

# 2 Neural non-ideal factors and Diversity of brain

## 2.1 Neuronal Inconsistency and Mismatch

### 2.1.1 Morphological Heterogeneity

Neurons, the fundamental units of the nervous system, exhibit an extraordinary 3D degree of morphological diversity that defies any notion of a uniform cellular blueprint [2.1]. This heterogeneity is not a mere biological curiosity but a foundational element of neural computation. The structure of a neuron—its soma (cell body) size, dendritic arborization complexity, axonal projection length, and synaptic bouton density—is intricately linked to its functional role within a circuit [2.2]. For instance, pyramidal neurons in the cerebral cortex possess elaborate apical dendrites that extend toward the pial surface, allowing them to integrate thousands of inputs from diverse sources across different cortical layers [2.3]. In contrast, cerebellar granule cells are among the smallest neurons, with short, bushy dendrites that receive input from a single mossy fiber, reflecting their role as high-throughput pattern separators [2.4]. This variation in scale is staggering; while the human brain contains approximately 86 billion neurons, the number of synapses exceeds 10^14, meaning each neuron can form tens of thousands of connections, but the specific morphology dictates how these connections are organized [2.5].

The branching patterns of dendrites alone display immense variability. Dendritic arbors range from the relatively simple bipolar shape of retinal ganglion cells to the highly complex, fractal-like trees of Purkinje cells in the cerebellum [2.6]. A single Purkinje cell's dendritic tree can span over half a millimeter and contain more than 100,000 synaptic spines [2.7]. This morphological complexity directly influences the neuron's computational power by determining its capacity for spatial summation and compartmentalized processing [2.8]. Different branches can function as semi-independent computational units, capable of local spike initiation, a phenomenon known as dendritic spikes [2.9]. Such structural features enable a single neuron to perform sophisticated computations, such as feature detection or coincidence detection, far beyond a simple threshold device.

This diversity extends to the subcellular level, including variations in organelle distribution and cytoskeletal architecture [2.10]. Mitochondria, critical for energy production, are often clustered near active synapses and nodes of Ranvier in myelinated axons, highlighting the tight coupling between energy demand and neuronal structure [2.11]. The endoplasmic reticulum also forms specialized structures like spine apparatuses in large dendritic spines, which regulate calcium dynamics crucial for synaptic plasticity [2.12]. The very existence of this profound morphological heterogeneity presents a significant challenge to neuroanatomists and modelers alike [2.13]. It suggests that the brain's information processing power arises not from identical components but from a vast library of uniquely shaped and connected processors, each contributing a specific function to the emergent properties of neural networks.

**2.1.2 Electrophysiological Variability**

Beyond their diverse shapes, neurons exhibit a remarkable spectrum of electrophysiological behaviors, forming what could be described as a "neural alphabet" of intrinsic firing patterns [2.14]. This variability is governed by the unique complement of voltage-gated ion channels expressed in each neuron's membrane, which act as molecular transistors to generate and modulate electrical signals [2.15]. Some neurons fire in regular, clock-like patterns of action potentials, known as tonic firing, exemplified by thalamocortical relay neurons that faithfully transmit sensory information [2.16]. Others display burst firing, where a rapid sequence of action potentials is followed by a period of silence, a behavior seen in certain midbrain dopamine neurons that is thought to encode reward prediction errors with high salience [2.17].

A key measure of electrophysiological identity is the neuron's response to injected current [2.18]. When presented with a step of depolarizing current, some neurons will fire immediately and then adapt their firing rate over time, while others may delay their first spike (delayed excitation) or show little adaptation at all [2.19]. Hippocampal CA1 pyramidal neurons, for example, are known for their pronounced spike-frequency adaptation, mediated by calcium-activated potassium channels, which allows them to respond strongly to new stimuli but filter out sustained inputs [2.20]. In contrast, fast-spiking interneurons, primarily GABAergic, express high levels of Kv3-family potassium channels that enable them to repolarize extremely quickly, allowing them to sustain firing rates exceeding 100 Hz without failure [2.21]. This makes them ideal for generating the precise, high-frequency oscillations observed in gamma rhythms, which are associated with cognitive processes like attention and memory binding [2.22].

This intrinsic diversity has profound implications for network dynamics [2.23]. It means that the same synaptic input will produce vastly different output patterns depending on the target neuron's biophysical properties [2.24]. A transient excitatory input might trigger a single spike in one cell but a prolonged burst in another [2.25]. Furthermore, the presence of subthreshold oscillations, resonance frequencies, and hysteresis in some neurons adds another layer of complexity, enabling them to act as band-pass filters or bistable switches [2.26]. This electrophysiological richness ensures that neural circuits are not homogeneous signal amplifiers but complex dynamical systems capable of rich, context-dependent responses [2.27]. The precise tuning of these ion channel ensembles during development and through experience is a central question in neuroscience, as it underpins the stability and adaptability of neural function [2.28]. Understanding this variability is essential for developing accurate models of brain function and for designing neuromorphic hardware that seeks to emulate neural computation [2.29].

### 2.1.3 Receptor and Neurotransmitter Heterogeneity

The communication between neurons is mediated by a vast and heterogeneous array of neurotransmitters and their corresponding receptors, creating a complex chemical language that far exceeds a simple binary code [2.30]. While glutamate and GABA serve as the primary excitatory and inhibitory neurotransmitters, respectively, their effects are modulated by a dizzying variety of receptor subtypes, each with distinct pharmacology, kinetics, and intracellular signaling cascades [2.31]. For example, glutamate acts on both ionotropic receptors (AMPA, NMDA, kainate), which are ligand-gated ion channels, and metabotropic receptors (mGluRs), which are G-protein-coupled receptors (GPCRs) [2.32]. The NMDA receptor, critical for Hebbian learning, is unique in its voltage-dependent magnesium block and its requirement for both ligand binding and postsynaptic depolarization to open, making it a perfect coincidence detector [2.33]. Similarly, GABA-A receptors are chloride channels whose activity can be potentiated by benzodiazepines and barbiturates, while GABA-B receptors are GPCRs that inhibit adenylyl cyclase and activate potassium channels, leading to slower, longer-lasting inhibition [2.34].

Beyond the main workhorses, the brain utilizes numerous neuromodulatory systems that use neurotransmitters like dopamine, serotonin, acetylcholine, and norepinephrine [2.35]. These systems typically project diffusely from subcortical nuclei and do not mediate point-to-point transmission but instead modulate the overall state of neural circuits [2.36]. Dopamine, for instance, plays a crucial role in reinforcement learning and motivation, acting on at least five distinct receptor types (D1-D5) that are coupled to different G-proteins, leading to either excitatory or inhibitory effects depending on the subtype and brain region [2.37]. This creates a landscape of opposing forces; for example, in the striatum, D1 receptors promote the direct pathway facilitating movement, while D2 receptors inhibit the indirect pathway suppressing unwanted movements [2.38]. Serotonin, with its 14 known receptor subtypes, is involved in mood, sleep, and appetite regulation, and its widespread influence contributes to the complex actions of many psychiatric drugs [2.39].

This chemical heterogeneity allows for fine-tuning of neural circuits on multiple timescales [2.40]. Fast ionotropic receptors mediate rapid synaptic transmission on the order of milliseconds, while metabotropic receptors and neuromodulators can alter neuronal excitability, synaptic strength, and gene expression over seconds to hours [2.41]. The co-expression of multiple receptor subtypes on a single neuron enables it to integrate signals from various neurotransmitter systems simultaneously, creating a highly nuanced response profile [2.42]. This complexity provides a rich substrate for pharmacological intervention but also makes it incredibly challenging to predict the net effect of a drug, as its action on one receptor subtype may be counteracted by its action on another [2.43]. The selective vulnerability of specific neuronal populations in disease, such as dopaminergic neurons in Parkinson's disease, underscores the critical importance of this receptor-specific architecture [2.44].

**2.1.4 Idiosyncratic Connectivity Patterns**

The connectome—the comprehensive map of neural connections in the brain—is defined by its inherent idiosyncrasy and lack of universal symmetry [2.45]. Each neuron's connectivity is shaped by a combination of spatiotemporally non-uniform 3D genetic program, developmental history, and experiential refinement, resulting in a unique wiring diagram for every individual and even between the two hemispheres of a single brain [2.46]. Unlike engineered systems built from standardized parts, the brain develops through a process of exuberant growth followed by competitive pruning [2.47]. During early development, neurons form far more connections than are ultimately needed. Synapses that are frequently co-activated are strengthened and preserved through Hebbian mechanisms ("fire together, wire together"), while inactive ones are eliminated [2.48]. This results in a final circuit that is highly efficient and optimized for the organism's environment, but one that is also inherently variable [2.49].

This variability manifests in several ways [2.50]. First, there is significant stochasticity in axon pathfinding and synapse formation. Even genetically identical individuals raised in similar environments will have subtly different neural circuits [2.51]. Second, the specificity of connections is often probabilistic rather than deterministic [2.52]. For example, a given pyramidal neuron in layer 5 of the visual cortex may project to multiple areas, including higher visual areas, the superior colliculus, and motor regions, but the exact number and strength of these projections vary between cells [2.53]. Third, recurrent networks, particularly prevalent in the neocortex and hippocampus, are characterized by dense, feedback loops where neurons are heavily interconnected with their neighbors [2.54]. This recurrent connectivity is believed to support functions like working memory, decision-making, and predictive coding, but it also introduces a high degree of sensitivity to initial conditions and potential for chaotic dynamics [2.55].

The consequences of this idiosyncratic wiring are profound [2.56]. It means that no two brains are wired exactly alike, providing a biological basis for individual differences in cognition, personality, and susceptibility to mental illness [2.57]. This variability is also a source of robustness; the loss of a single connection in a redundant, distributed network is less likely to cause catastrophic failure than in a serially wired system [2.58]. However, it poses a monumental challenge for neuroscience. Mapping the complete connectome of a mammalian brain, with its hundreds of billions of neurons and quadrillions of synapses, remains an unsolved problem [2.59]. Current technologies like electron microscopy can reconstruct small volumes, but scaling up to whole brains is computationally prohibitive [2.60]. Thus, the field often relies on population-level averages, which can obscure the critical functional role of rare or unique connectivity motifs [2.61].

**2.1.5 Functional Consequences of Neuronal Mismatch**

The cumulative effect of neuronal inconsistency—spanning morphology, electrophysiology, chemistry, and connectivity—is not a flaw to be corrected but a core feature that confers immense functional advantages to the nervous system [2.62]. This mismatch ensures that neural processing is inherently parallel, redundant, and resilient [2.63]. By distributing a computational task across a population of neurons with slightly different response properties, the brain can achieve robust representations that are immune to the failure of any single unit [2.64]. This principle is elegantly demonstrated in the visual system, where the perception of a simple edge is encoded not by a single "edge detector" neuron but by a population of neurons in V1, each tuned to a slightly different orientation [2.65]. The collective activity of this population provides a more reliable and noise-resistant representation than a single neuron ever could [2.66].

Furthermore, this diversity is the foundation of distributed coding schemes, which are far more powerful and flexible than localist codes [2.67]. In a localist scheme, one neuron equals one concept, which is highly vulnerable to damage [2.68]. In a distributed scheme, a concept is represented by a specific pattern of activity across a large ensemble of neurons [2.69]. Because the neurons in this ensemble are heterogeneous, they each contribute a unique facet to the representation, making the code exponentially richer and more robust [2.70]. This heterogeneity also underpins the brain's ability to generalize [2.71]. A network composed of identical units would struggle to extrapolate beyond its training data, whereas a diverse network can leverage its varied response profiles to make predictions about novel situations [2.72]. For instance, when encountering a new animal, the brain can draw upon a distributed representation built from experiences with cats, dogs, and foxes, synthesizing a prediction based on shared features, despite no single neuron being an expert on "new animals" [2.73].

This neuronal mismatch is also a wellspring of creativity and behavioral flexibility [2.74]. As discussed in the context of creativity and dreams, the "noise" generated by this diversity allows for information roaming and entanglement, enabling the brain to explore novel combinations of ideas [2.75]. The fact that neural circuits are not perfectly optimized for efficiency alone, but retain a degree of redundancy and stochasticity, allows for exploratory behaviors and adaptive solutions to unforeseen problems [2.76]. In essence, the brain's apparent "messiness" is its greatest strength, transforming a collection of unreliable biological components into a highly intelligent, adaptive, and creative system capable of navigating an infinitely complex world [2.77]. This perspective reframes neuronal inconsistency from a source of error to a fundamental principle of intelligent design in biology [2.78].

## 2.2 Cellular and Molecular Sources of Neural Noise

## 2.2.1 Stochastic Fluctuations in Synaptic Neurotransmitter Release

At the heart of neural communication lies the synapse, a site of profound molecular stochasticity that fundamentally limits the reliability of information transfer [2.79]. One of the most significant sources of neural noise is the probabilistic nature of neurotransmitter release. When an action potential arrives at the presynaptic terminal, it triggers the opening of voltage-gated calcium channels, leading to a local influx of $Ca^{2+}$ ions. This calcium surge binds to sensor proteins, primarily synaptotagmin, which then catalyzes the fusion of synaptic vesicles with the presynaptic membrane, releasing their contents into the synaptic cleft [2.80]. However, this process is inherently unreliable. Not every action potential leads to the release of a quantum of neurotransmitter; the probability of release (Pr) can vary widely, from less than 0.1 to nearly 1.0, depending on the synapse type, recent activity history (due to facilitation or depression), and neuromodulatory state [2.81].

This randomness arises from the low number of molecules involved in the process. A single synaptic vesicle contains only a finite number of neurotransmitter molecules (typically thousands), and the fusion event itself is a thermally driven process subject to Brownian motion [2.82]. Even if a vesicle is docked and primed, the conformational changes in the SNARE complex proteins required for membrane fusion are stochastic events [2.83]. Consequently, repeated presentations of the exact same stimulus can result in wildly different postsynaptic responses—a phenomenon known as synaptic variability [2.84]. For example, stimulating a presynaptic neuron at a constant frequency might produce a barrage of postsynaptic potentials in one trial and fail to elicit any detectable response in the next. This noise is not merely a nuisance; it is a defining characteristic of biological synapses that artificial systems, which rely on deterministic digital logic, completely lack.

The functional impact of this stochastic release is dual-edged [2.85]. On one hand, it introduces a fundamental limit on the precision of neural coding, making it difficult for downstream neurons to decode subtle variations in input timing or rate. On the other hand, this randomness can be harnessed by the brain for beneficial purposes. It serves as a source of internal variability, preventing neural circuits from becoming trapped in rigid, maladaptive states. In learning algorithms, a controlled amount of noise can aid exploration in a search space, helping to avoid local minima. Similarly, in the brain, synaptic noise may facilitate probabilistic decision-making and behavioral exploration. For instance, in the absence of clear sensory cues, the random fluctuations in synaptic transmission could nudge an animal to try a new foraging path, potentially leading to a discovery. Thus, the very imperfection of the synaptic transmission mechanism may be a key ingredient in the brain's ability to adapt and innovate.

**2.2.2 Membrane Potential Instability and Ion Channel Noise**

The resting membrane potential of a neuron is not a static value but a dynamic equilibrium maintained against a constant barrage of molecular noise [2.86]. A primary contributor to this instability is ion channel noise, arising from the discrete and stochastic gating of the millions of ion channels embedded in the neuronal membrane [2.87]. Each channel exists in a limited number of conformational states: closed, open, and sometimes inactivated. The transition between these states is driven by thermal energy and is fundamentally probabilistic. For any given set of conditions—such as membrane voltage and ligand concentration—there is only a certain probability that a channel will be open at any instant.

The conductance of the membrane is determined by the sum of all open channels. Because the number of channels is finite, the total conductance fluctuates randomly around its mean value [2.88]. This is analogous to flipping a coin: the more coins you flip (or the more channels you have), the closer the proportion of heads (open channels) will be to 50% (the expected probability), according to the Principle of large numbers. However, with fewer channels, the relative fluctuations (the "noise") become larger. This is particularly relevant for small cellular compartments like thin dendrites or axon terminals, where the number of channels is lower, leading to greater relative conductance fluctuations [2.89]. These fluctuations manifest as random, miniature currents across the membrane, causing the membrane potential to jitter constantly, even in the absence of synaptic input.

This intrinsic noise has significant consequences for neuronal excitability. It lowers the effective threshold for action potential generation. Instead of requiring a precise amount of depolarizing current to reach a fixed threshold, a neuron may fire earlier or later due to a fortuitous alignment of open sodium channels that provide an extra boost of inward current. Conversely, a cluster of open potassium channels could hyperpolarize the membrane and prevent firing even if the average drive is sufficient. This phenomenon, known as stochastic resonance, can paradoxically enhance the detection of weak signals [2.90]. A sub-threshold signal combined with the right amount of noise can occasionally push the membrane potential over the firing threshold, allowing the signal to be detected. Without the noise, the signal would remain invisible. This demonstrates that the brain does not strive for silent, noise-free operation but operates in a regime where noise and signal interact synergistically to optimize performance in uncertain environments.

**2.2.3 Variability in Action Potential Thresholds**

The action potential, or spike, is the fundamental unit of neural communication, yet its generation is subject to surprising variability [2.91]. Contrary to the textbook simplification of a fixed voltage threshold, the threshold for spike initiation is dynamic and influenced by numerous factors, adding another layer of noise to neural signaling. Traditionally, the threshold was considered a stable property of a neuron, perhaps around -55 mV. However, modern intracellular recordings have shown that this value can shift by several millivolt depending on the neuron's immediate history [2.92]. After a series of action potentials, the threshold often becomes more negative (lower), making the neuron more excitable—a phenomenon related to the slow recovery of sodium channels from inactivation. Conversely, prior hyperpolarization can lead to a higher threshold.

This variability is primarily attributed to the state-dependent kinetics of voltage-gated sodium (NaV) channels [2.93]. Following an action potential, a fraction of NaV channels enter a slow-inactivated state and require time to recover before they can open again. If a subsequent depolarizing input arrives during this recovery period, fewer NaV channels are available to generate the regenerative inward current needed for a spike, effectively raising the threshold. Additionally, the activation and inactivation curves of these channels are themselves sensitive to temperature and the composition of the surrounding lipid bilayer, introducing further biological variability. The location of spike initiation also plays a crucial role. In many neurons, spikes are initiated not at the soma but in the axon initial segment (AIS), a specialized region enriched with NaV channels [2.94]. The AIS can undergo activity-dependent plasticity, changing its length or channel density, thereby altering the overall excitability and threshold of the neuron.

This dynamic threshold has important functional implications. It transforms the neuron from a simple integrator of inputs into a more complex adaptive filter. The neuron's responsiveness is not fixed but is modulated by its recent activity, allowing it to adjust its gain based on the prevailing input statistics. This is a key mechanism for maintaining homeostasis and preventing runaway excitation or silencing in neural circuits. It also means that the same synaptic input can produce different outcomes at different times, contributing to the temporal unpredictability of neural responses. For an external observer, this appears as noise, but for the brain, it represents a sophisticated mechanism for regulating its own sensitivity and ensuring stable operation in a noisy, fluctuating world.

**2.2.4 Thermal Noise in Neural Signaling**

At the most fundamental physical level, neural signaling is constrained by the principles of thermodynamics, and thermal noise is an unavoidable consequence [2.95]. This refers to the random, incessant motion of atoms and molecules—Brownian motion—that occurs at any temperature above absolute zero. In the context of neurons, thermal noise affects virtually every stage of signal transduction, from the opening of ion channels to the diffusion of neurotransmitters and second messengers [2.96]. At physiological temperatures (~37°C for humans), the kinetic energy of molecules is substantial, causing constant collisions and vibrations that perturb the delicate molecular machinery of the cell.

For ion channels, thermal noise is the driving force behind the stochastic transitions between their open and closed states [2.97]. The energy barrier separating these states is on the same order of magnitude as the thermal energy ($k_BT$), making the transitions highly susceptible to thermal agitation. Similarly, the diffusion of neurotransmitters across the synaptic cleft is a thermally driven process, leading to variability in the time it takes for a molecule to reach its receptor and in the number that arrive within a given timeframe. Once bound, the receptor-ligand interaction itself is a dynamic equilibrium, with the neurotransmitter dissociating and reassociating due to thermal jostling.

While this noise is omnipresent, its impact is generally small compared to other sources of biological noise, such as synaptic release failure or ion channel flickering. However, it sets a fundamental lower bound on the fidelity of neural communication. In electronic circuits, engineers combat thermal noise (Johnson-Nyquist noise) by increasing signal amplitude or using cooling. The brain, however, operates at room temperature and cannot simply increase the "signal" (i.e., the number of ions or molecules) without incurring massive metabolic costs [2.98]. Instead, the brain has evolved strategies to operate reliably despite this physical limitation. One such strategy is redundancy: by averaging the activity of many independent, noisy elements (neurons or synapses), the overall signal-to-noise ratio can be improved [2.99]. Another is the use of positive feedback loops, such as the regenerative opening of NaV channels during an action potential, which rapidly amplifies a small depolarizing signal into a full-blown spike, effectively overcoming the background thermal fluctuations. This interplay between physical constraints and biological evolution highlights the elegance with which the brain has solved the engineering challenge of building a reliable computing machine from inherently noisy components.

**2.2.5 Glial-Neuronal Mismatch and Its Impact on Homeostasis**

The traditional view of the brain as a neuron-centric organ is increasingly being replaced by a more holistic understanding of the tripartite synapse, where astrocytes and other glial cells play an indispensable and dynamic role in regulating neuronal function and maintaining homeostasis [2.100]. Astrocytes, star-shaped glial cells, ensheath synapses with their fine processes, creating a microenvironment that profoundly influences synaptic transmission and plasticity [2.101]. They perform critical tasks such as clearing neurotransmitters (like glutamate) from the synaptic cleft via high-affinity transporters [2.102], supplying metabolic substrates (like lactate) to fuel neuronal activity [2.103], buffering extracellular $K^+$ ions to maintain ionic balance, and modulating blood flow to match local metabolic demands (neurovascular coupling).

A key source of non-ideal factors arises from the asynchronous and complex signaling between neurons and glia [2.104]. Unlike fast synaptic transmission, glial responses are typically much slower, occurring on timescales of hundreds of milliseconds to minutes. An astrocyte may detect a surge in neuronal activity through uptake of $K^+$ or neurotransmitter spillover, leading to an increase in intracellular $Ca^{2+}$ concentration [2.105]. This $Ca^{2+}$ wave can propagate through the astrocytic network and trigger the release of gliotransmitters, such as ATP or D-serine, which can then feed back onto nearby neurons to modulate synaptic strength. This feedback loop is indirect, diffuse, and lacks the precise spatial and temporal resolution of neuronal signaling. It is not targeted to a single synapse but can affect a neighborhood of synapses, potentially inducing widespread changes in network excitability.

This mismatch between the fast, point-to-point communication of neurons and the slower, more diffuse modulation by glia is not a bug but a feature [2.106]. It allows glia to integrate information over time and space, serving as a thermostat for neural circuits. For example, during prolonged periods of high activity, astrocytic buffering and metabolic support prevent excitotoxicity and energy depletion. However, this system is vulnerable to dysfunction. In pathological conditions like epilepsy or Alzheimer's disease, the astrocyte's ability to buffer $K^+$ or clear glutamate can be impaired, leading to a breakdown in homeostasis and runaway neuronal excitation [2.107]. Furthermore, the heterogeneity of astrocyte populations themselves adds another layer of complexity; astrocytes in different brain regions or even within the same region can have distinct molecular profiles and functional capabilities, leading to regional differences in how homeostasis is maintained. This intricate, multi-timescale dialogue between neurons and glia is a fundamental source of non-ideal factors that shapes the dynamic stability of the entire nervous system.

### 2.2.6 Metabolic Noise

Neural computation is an energetically expensive process, consuming a disproportionate amount of the body's resources [2.108]. This intense metabolic demand introduces a significant source of noise and variability into neural function. Neurons rely almost exclusively on aerobic glycolysis and oxidative phosphorylation to generate ATP, which powers essential processes like maintaining the resting membrane potential via the $Na^+/K^+$ ATPase pump, recycling neurotransmitters, and supporting active transport along axons and dendrites. The supply of glucose and oxygen is delivered by the bloodstream, but this delivery is not constant. Blood flow fluctuates due to cardiac pulsations, changes in posture, respiration, and autonomic control, leading to minute-by-minute variations in the availability of metabolic substrates [2.109].

This variability in energy supply translates directly into metabolic noise at the cellular level [2.110]. A neuron experiencing a transient dip in oxygen or glucose may have reduced ATP production, impairing the function of ATP-dependent pumps and enzymes. This can lead to a partial collapse of ionic gradients, making the neuron more excitable and prone to aberrant firing. Conversely, a surge in energy supply might temporarily enhance the neuron's ability to sustain high-frequency firing. This metabolic noise is particularly relevant in the context of functional imaging techniques like fMRI, which measures the blood-oxygen-level-dependent (BOLD) signal as a proxy for neural activity [2.111]. The BOLD signal reflects this dynamic interplay between neural activity and vascular supply, but it is delayed and smoothed compared to the underlying electrical activity, adding a layer of temporal uncertainty.

Moreover, the allocation of energy within a neuron is a strategic decision. Under metabolic stress, a neuron may prioritize essential functions, such as maintaining basic excitability, at the expense of higher-order processes like long-term synaptic plasticity [2.112]. This can introduce a bias into learning rules, favoring the strengthening of already strong synapses (which are more metabolically efficient) over the formation of new ones. This link between metabolism and plasticity is highlighted by research showing that mitochondrial function and biogenesis are tightly coupled to learning [2.113]. For example, studies have shown that growth hormone is required for the maturation of hippocampal memory engram cells, linking protein synthesis and metabolic health directly to the consolidation of long-term memories. Thus, the brain's function is not insulated from the body's physiological state; metabolic noise is a pervasive factor that links the internal milieu to the very fabric of neural computation and learning.

### 2.2.7 Protein Aggregation Variability among Neurons

The integrity of neuronal function depends critically on the proper folding and degradation of proteins. A major source of non-ideal factors and a hallmark of aging and neurodegenerative diseases is the accumulation of misfolded and aggregated proteins 2-114

. The proteostasis network, comprising chaperones, the ubiquitin-proteasome system (UPS), and autophagy pathways, works tirelessly to maintain protein quality control. However, this system is imperfect and exhibits significant variability between neurons. Some neurons appear to be more resilient to proteotoxic stress, while others are selectively vulnerable to the accumulation of specific aggregates, such as amyloid-beta (in Alzheimer's), tau (in Alzheimer's and frontotemporal dementia), or alpha-synuclein (in Parkinson's) [2.115].

This differential vulnerability can be attributed to several factors. First, different neuron types have varying basal metabolic rates and levels of oxidative stress, which can damage proteins and overwhelm repair mechanisms [2.116]. For instance, dopaminergic neurons in the substantia nigra are particularly vulnerable because dopamine metabolism itself generates reactive oxygen species. Second, neurons differ in their expression levels of protective chaperone proteins and the efficiency of their lysosomal degradation pathways. Third, the morphology of the neuron plays a crucial role. Neurons with exceptionally long axons, like motor neurons, face a significant logistical challenge in transporting damaged proteins back to the soma for degradation, making distal regions prone to the buildup of aggregates [2.117].

The stochastic nature of protein aggregation adds another layer of noise [2.118]. Misfolding is a random event, and once a seed of an aggregate forms, it can grow and spread in a prion-like fashion, templating the misfolding of native proteins. This process is inherently unpredictable, explaining why disease progression can vary so dramatically between individuals, even with similar genetic predispositions. This variability is not just pathological; even in healthy brains, low levels of protein aggregation occur and may contribute to normal age-related cognitive decline [2.119]. From a systems perspective, this ongoing battle against entropy and decay is a fundamental constraint on neural longevity. The need to continuously monitor and repair protein infrastructure consumes valuable metabolic resources and imposes a ceiling on the theoretical lifespan of a neural circuit. This persistent threat of proteostatic failure is a sobering reminder of the fragile, maintenance-intensive nature of the biological brain.

## 2.3 Circuit-Level non-ideal factors Arising from Heterogeneous Components
### 2.3.1 Oscillatory Desynchronization Due to Neuronal Diversity

Oscillations, rhythmic fluctuations in the electrical activity of neuronal populations, are a ubiquitous feature of brain dynamics, observed across various frequency bands such as delta (1–4 Hz), theta (4–8 Hz), alpha (8–12 Hz), beta (12–30 Hz), and gamma (>30 Hz) [2.120]. These oscillations are thought to play a crucial role in organizing neural communication, facilitating processes like attention, memory formation, and sensory binding through a mechanism called "communication through coherence" [2.121]. According to this hypothesis, two neural groups can communicate efficiently only if their oscillatory phases are aligned, allowing their respective windows of excitability to coincide. However, the very diversity that gives rise to functional richness also creates a fundamental challenge to the synchronization of these rhythms, leading to a state of perpetual desynchronization.

This desynchronization is primarily driven by the heterogeneity of the constituent neurons. Consider a population of neurons that should ideally fire in unison to generate a coherent gamma rhythm. In reality, these neurons will have slight differences in their intrinsic firing properties (e.g., due to variations in ion channel density as discussed in 2.1.2), different synaptic input histories, and varying degrees of neuromodulatory tone. These small differences cause each neuron to advance or lag slightly in its cycle, accumulating over time to create phase dispersion within the population [2.122]. A neuron with a slightly faster intrinsic oscillation will gradually pull ahead of its peers, while one with a higher spike threshold will fall behind. This results in a broadened phase distribution, weakening the overall power of the oscillation and reducing its effectiveness as a timing signal.

Despite this tendency toward desynchronization, the brain has evolved mechanisms to harness it. Desynchronization is not always detrimental; it can serve as a functional switch. For instance, a transition from a synchronized state (associated with focused attention) to a desynchronized state (associated with arousal and sensory processing) is a common feature of wakefulness. More importantly, this constant tug-of-war between synchronizing forces (like recurrent excitatory connections and reciprocal inhibition) and desynchronizing forces (neuronal diversity) keeps neural networks in a dynamic, metastable state [2.123]. This prevents the system from locking into rigid, fixed-point attractors and allows for flexible switching between different cognitive states. Thus, the "noise" introduced by neuronal diversity ensures that the brain's oscillatory dynamics are not like a perfect metronome but more like a jazz ensemble, where slight improvisations and variations maintain the music's vitality and prevent it from becoming monotonous.

**2.3.2 Feedback Delay Mismatches in Recurrent Networks**

Recurrent neural networks, which are characterized by extensive feedback loops where outputs are fed back as inputs, are a dominant architectural motif in the brain, particularly in the cortex and hippocampus [2.124]. These networks are powerful because they can maintain internal states, allowing for functions like working memory, sequential processing, and predictive modeling. However, this power comes at a cost: they are inherently unstable and highly sensitive to timing delays. The introduction of a delay in a feedback loop can transform a stabilizing negative feedback signal into a destabilizing positive feedback signal, leading to oscillations, runaway excitation, or system failure [2.125]. In an idealized system, all components would have identical and predictable delays. In the biological brain, however, this is far from the case.

Feedback delay mismatches arise from several sources. First, the propagation speed of action potentials along axons varies significantly. Axon diameter, myelination status, and even temperature all influence conduction velocity [2.126]. A thick, myelinated axon can conduct signals at over 100 meters per second, while a thin, unmyelinated one may travel at less than 1 meter per second. In a recurrent loop connecting distant brain regions, the feedback signal may take tens or even hundreds of milliseconds to return. If the network's dynamics change on a shorter timescale than this delay, the feedback will be based on outdated information and can be inappropriate. Second, synaptic transmission itself introduces delays. Fast ionotropic synapses have latencies of 1-2 ms, but slower metabotropic synapses can add tens of milliseconds of delay [2.127]. Neuromodulatory systems, which act on much longer timescales (seconds to minutes), represent an extreme form of delayed feedback.

These delays create a significant non-ideal factors for real-time computation. A classic example is the vestibulo-ocular reflex (VOR), which stabilizes vision during head movement. The sensory input (head motion) must be processed and transformed into a motor command (eye movement) with minimal delay. To compensate for inevitable neural delays, the brain uses predictive, internal models [2.128]. It doesn't just react to current head position; it estimates future head motion and commands the eye to move preemptively. This requires the network to be calibrated to the specific delays of its own circuitry. Any mismatch between the actual delay and the predicted delay (for instance, due to fatigue or pathology) can lead to a poorly tuned reflex and blurred vision. The necessity for such complex predictive mechanisms is a direct consequence of the brain's physical embodiment and the inherent non-ideal timing of its biological components.

**2.3.3 Cross-Modal Crosstalk Mediated by Atypical Connectivity**

The brain is traditionally divided into specialized modules for processing different sensory modalities—visual cortex, auditory cortex, somatosensory cortex, etc. However, this segregation is not absolute, and "atypical connectivity" in the form of cross-modal connections is a fundamental source of non-ideal factors that blurs these boundaries [2.129]. These connections allow for the integration of information from different senses, which is essential for creating a unified and coherent percept of the world. Yet, they also create pathways for unintended "crosstalk," where activity in one sensory system can inadvertently influence processing in another.

One prominent example of beneficial crosstalk is multi-sensory integration. Visual input can enhance the detection of faint sounds (e.g., seeing someone's lips move makes their voice easier to hear in a noisy room), and tactile sensations can influence visual perception (e.g., feeling a texture can change how it looks) [2.130]. This integration occurs in specialized association areas, like the superior temporal sulcus, which receive convergent input from multiple sensory cortices. However, the connectivity is often reciprocal and not perfectly gated, meaning that under certain conditions, such as inattention or altered states of consciousness, these normally helpful interactions can become maladaptive. For instance, in conditions like Charles Bonnet syndrome, damage to the visual system can lead to vivid hallucinations, possibly due to unmasked cross-talk from other sensory areas or top-down generative processes running unchecked.

Another example is synesthesia, a condition where stimulation of one sensory pathway leads to automatic, involuntary experiences in a second pathway, such as seeing colors when hearing music [2.131]. While once considered a rare anomaly, research suggests that the neural substrates for synesthetic experiences exist in all brains but are usually suppressed. This implies that the typical adult brain has developed inhibitory mechanisms to prevent excessive crosstalk and maintain functional segregation. The persistence of some atypical connections, even if inhibited, is a form of non-ideal factors that can be revealed under specific circumstances, such as after taking psychedelics, which reduce top-down control [2.132]. This latent connectivity might be a vestige of early development, where the brain starts with an overabundance of connections that are later pruned. The fact that this atypical connectivity exists at all, even if suppressed, is a testament to the brain's inherent complexity and its departure from a modular, "clean-room" design, instead favoring a rich, interconnected web of associations that supports creativity and multimodal understanding.

## 2.4 Cognitive and Psychological Manifestations of Neural Non-Ideal Factors

### 2.4.1 Individual Learning Biases Rooted in Neuronal Variability

Individual differences in learning styles and abilities, often perceived as matters of personal preference or effort, have deep roots in the biological ideal factors of the brain [2.133]. The morphological, electrophysiological, and chemical heterogeneity of neurons creates a unique neural architecture in every individual, leading to inherent biases in how information is acquired, processed, and retained. For example, the density and configuration of dendritic spines in the prefrontal cortex, a region critical for executive function and working memory, can vary significantly between people [2.134]. Individuals with a higher density of spines on certain pyramidal neurons may have a greater capacity for holding and manipulating information online, giving them a natural advantage in learning complex, rule-based subjects like mathematics or programming.

Similarly, variability in neurotransmitter systems can lead to different learning motivations and reward sensitivities. The dopaminergic system, which mediates reward prediction error and reinforces successful behaviors, shows considerable individual variation [2.135]. Some individuals have a more sensitive reward system, making them highly motivated by positive feedback and quick to learn from successes. Others may be more sensitive to losses or punishments, adopting a risk-averse learning style. This is reflected in the "learning curve" and can be observed in educational settings. A student who thrives on competition and public recognition may have a dopamine system tuned to social rewards, while a student who prefers solitary study and avoids tests may have a system more sensitive to the aversive aspects of evaluation.

Furthermore, the development of the brain's self-regulation systems, such as those described in the theory of self-regulated learning (SRL), is a direct product of how these neural systems mature [2.136]. SRL involves goal setting, strategy planning, progress monitoring, and reflection. The ability to focus attention, a key component of SRL, is dependent on the development of neural circuits involving the prefrontal cortex and its connections to other brain regions. A child with a naturally more distractible attention system may find it harder to engage in deep, focused work, not due to a lack of intelligence, but because of a neurobiological difference. Over time, these initial biases can be amplified by experience; a child who finds reading easy may read more, strengthening their linguistic neural networks, while a child who struggles may avoid reading, falling further behind. Thus, the brain's non-ideal architecture establishes a starting point for a lifelong trajectory of learning, where biology and experience interact in a continuous feedback loop to shape cognitive profiles.

**2.4.2 Emotional Response Mismatch across Individuals**

The profound individual differences in emotional experience—from baseline temperament to reactivity in stressful situations—are a direct manifestation of neural non-ideal factors at the circuit level [2.137]. The limbic system, a network of structures including the amygdala, hippocampus, hypothalamus, and prefrontal cortex, orchestrates emotional responses, but the way these structures are wired and regulated varies greatly from person to person. The amygdala, for instance, is central to processing fear and threat. Research has shown that the volume and functional connectivity of the amygdala correlate with trait anxiety; individuals with a larger amygdala or stronger amygdala-prefrontal connectivity may be more vigilant and reactive to potential threats, interpreting ambiguous social cues as hostile [2.138]. This creates a fundamental mismatch in emotional experience: one person may perceive a minor critique as a devastating attack, while another may brush it off as constructive feedback.

Hormonal systems also contribute significantly to this variability. The hypothalamic-pituitary-adrenal (HPA) axis regulates the stress response through the release of cortisol. There is a wide normal range in HPA axis reactivity. Some individuals mount a large, rapid cortisol response to stress, which can be beneficial for acute challenges but damaging if chronically elevated. Others have a more dampened response [2.139]. This biological difference underlies the common observation that some people thrive under pressure, while others become paralyzed. Furthermore, the role of sex hormones like estrogen and testosterone in modulating mood and aggression adds another dimension of complexity, contributing to differences in emotional expression and disorders like depression and PTSD between genders [2.140].

This neural basis for emotional mismatch has important psychological and social consequences. It explains why therapeutic approaches that work for one person may be ineffective for another; a treatment targeting serotonergic pathways may help someone with a specific imbalance but not someone whose emotional dysregulation stems from a different neural circuit. It also highlights the importance of empathy and personalized understanding in social interactions. Recognizing that a colleague's irritability or a partner's sadness may stem from their unique neurobiology, rather than a character flaw, can foster more compassionate relationships. Ultimately, the "emotional fingerprint" of an individual, as unique as their DNA, is a tapestry woven from the non-ideal, heterogeneous threads of their neural circuitry, demonstrating that our inner lives are deeply rooted in the physical matter of our brains.

**2.5 Pathological Implications of Selective Neuronal Vulnerability**

**2.5.1 Differential Susceptibility of Neuron Subtypes to Degeneration**

Neurodegenerative diseases such as Alzheimer's, Parkinson's, and amyotrophic lateral sclerosis (ALS) are characterized by a striking selectivity in which specific types of neurons degenerate, while others in close proximity remain largely unaffected [2.141]. This selective neuronal vulnerability is one of the clearest demonstrations that the brain's non-ideal architecture has profound pathological consequences. The reasons for this selectivity are multifactorial, stemming from a convergence of intrinsic cellular properties and extrinsic environmental stresses. For instance, in Parkinson's disease, dopaminergic neurons in the substantia nigra pars compacta (SNpc) are lost, leading to the characteristic motor symptoms. These neurons are vulnerable because of their unique physiology: they have a slow, pacemaker-like activity that causes chronic elevation of intracellular calcium, placing a high metabolic demand on mitochondria [2.142]. This sustained stress, combined with the oxidative byproducts of dopamine metabolism, creates a perfect storm for mitochondrial dysfunction and the eventual accumulation of toxic alpha-synuclein aggregates.

In contrast, neighboring dopaminergic neurons in the ventral tegmental area (VTA) are relatively spared, despite having similar neurochemical identities. This difference is attributed to variations in their ion channel repertoire; VTA neurons use different calcium channels that are less taxing on mitochondria. In ALS, upper motor neurons in the motor cortex and lower motor neurons in the spinal cord are specifically targeted. These neurons are uniquely vulnerable due to their enormous size and the extreme metabolic burden of maintaining their exceptionally long axons, which can stretch from the brainstem to the toes [2.143]. This long-distance transport makes them highly dependent on efficient axonal transport and protein quality control systems, which are disrupted in ALS. The fact that other large neurons, like sensory neurons, are not affected points to additional, unknown factors, possibly related to specific patterns of synaptic input or gene expression.

This selective vulnerability pattern provides a crucial clue for researchers seeking to understand and treat these diseases. It suggests that the key to protection may lie not in treating the entire brain uniformly but in bolstering the specific weaknesses of vulnerable cell types [2.144]. For example, therapies aimed at reducing calcium load in SNpc neurons or enhancing mitochondrial health in motor neurons are being actively explored. Understanding why some neurons are resistant can also reveal powerful endogenous neuroprotective mechanisms. This knowledge is guiding the development of targeted gene therapies and precision medicine approaches, moving away from broad, blunt-force treatments toward interventions that address the specific "Achilles' heel" of the neurons most at risk.

**2.5.2 Emergence and Impact of Cancerous Neuron-Like Cells**

While true cancerous transformation of post-mitotic neurons is exceedingly rare because they do not divide, the broader category of brain tumors often arises from the supportive glial cells (gliomas) or from precursor cells [2.145]. The emergence of these cancerous, neuron-like cells represents a catastrophic failure of the brain's normal regulatory systems and a profound disruption of its non-ideal equilibrium. Glioblastoma multiforme (GBM), the most aggressive form of glioma, is notorious for its invasive growth and resistance to therapy. A key feature of GBM is its phenotypic plasticity, where tumor cells can adopt characteristics reminiscent of different neural cell types, including neurons. This "neuron-like" behavior is not benign mimicry; it allows the tumor cells to hijack the brain's normal signaling environment [2.146].

For example, research has shown that GBM cells can form functional synapses with host neurons, receiving glutamatergic input. This activity drives the proliferation of the tumor cells, creating a dangerous positive feedback loop where neural activity fuels cancer growth. This phenomenon, termed "tumor-nervous system coupling," is a stark illustration of how a pathological entity can exploit the existing neural architecture [2.147]. The cancerous cells are not constrained by the normal ideal factors of the brain; they can use its rich connectivity and signaling molecules for their own malignant purposes. This leads to severe neurological impacts, including seizures, which are a common symptom of brain tumors, caused by the tumor's disruption of normal neural circuit stability and induction of hyperexcitability.

The impact of these cancerous cells goes beyond mass effect and invasion. They secrete various factors that disrupt the blood-brain barrier, promote angiogenesis, and suppress the immune system, creating a permissive microenvironment for their survival. The metabolic demands of the growing tumor can starve surrounding healthy tissue, exacerbating neuronal dysfunction [2.148]. This creates a vicious cycle where the tumor-induced non-ideal factor—through abnormal signaling, metabolic competition, and inflammation—leads to progressive cognitive and motor decline. The study of these tumors not only aids in developing better oncological treatments but also provides unexpected insights into the normal function of neural circuits. By observing how the system breaks down when infiltrated by a parasitic, neuron-like agent, scientists can gain a deeper appreciation for the delicate balance and resilience of the healthy, albeit non-ideal, brain.

## 2.6 Massive Numbers of Neurons and Dendrites, Terminals, and Synapses Forming Non-uniform Brain Regions, Cortex, and Left-Right Hemispheres

### 2.6.1 Non-uniform Brain Regions and Diverse Regional Connections

The human brain is not a homogeneous lump of tissue but a mosaic of highly specialized, non-uniform regions, each with a distinct cellular composition, connectivity profile, and functional specialization [2.149]. This macroscopic non-uniformity is a direct consequence of the massive numbers of neurons and their complex connections. From the evolutionarily ancient brainstem, which controls vital functions like breathing and heart rate, to the photogenically newer neocortex, responsible for higher cognition, the brain is organized into a hierarchy of interacting subsystems. These regions are not isolated islands; they are densely interconnected through tracts of white matter, forming a complex network with a small-world topology—highly clustered locally yet efficiently connected globally [2.150].

This regional diversity is evident in their anatomy. The primary visual cortex (V1) is characterized by a distinctive striped appearance due to its columnar organization, while the hippocampus has a curved, seahorse-like shape with a laminated structure [2.151]. The functional roles of these regions are similarly divergent: the amygdala is central to emotion, the basal ganglia to motor control, and the prefrontal cortex to executive functions like planning and decision-making. The connections between them are equally diverse and specific. The thalamus acts as a central relay, routing sensory information from the periphery to the appropriate cortical areas [2.152]. The corpus callosum, the largest commissural tract, connects homologous regions of the two cerebral hemispheres, allowing for inter hemispheric communication. The strength and specificity of these connections determine the brain's functional architecture. For instance, the dorsal stream of visual processing ("where" pathway) projects from visual cortex to parietal cortex, enabling spatial navigation, while the ventral stream ("what" pathway) projects to temporal cortex, enabling object recognition [2.153].

This intricate web of regional connections creates a system of immense complexity. The number of possible connections between these regions scales combinatorially, leading to a virtually infinite number of potential network configurations. This structural richness is the foundation of the brain's functional repertoire. However, it also means that damage to a single region or a specific connection can have widespread and often unpredictable consequences. For example, a stroke affecting the left hemisphere's Broca's area can cause expressive aphasia, disrupting speech production, while leaving comprehension relatively intact [2.154]. This non-uniformity ensures that the brain is a robust and adaptable system, capable of complex, integrated behavior, but it also makes it highly vulnerable to focal insults.

### 2.6.2 Non-uniform Cerebral Cortex

The cerebral cortex, the outermost layer of the brain, is a paradigm of non-uniform structure and function [2.155]. Covering approximately 2 square feet in an adult human and folded into gyri and sulci to fit within the skull, it is the seat of conscious thought, perception, and voluntary movement. Its non-uniformity is organized on multiple levels. First, it is divided into four main lobes—frontal, parietal, temporal, and occipital—each with a general functional domain. Second, within these lobes, it is further subdivided into dozens of cytoarchitectonically distinct areas, originally mapped by Korbinian Brodmann based on differences in the six-layered structure of the cortex under a microscope [2.156].

Each cortical area has a unique arrangement of neurons, differing in layer thickness, cell density, and cell type proportions. For example, primary motor cortex (M1) has a very large layer V, containing giant Betz cells whose axons descend to the spinal cord to control muscles [2.157]. In contrast, primary visual cortex (V1) has a prominent layer IV, which receives the bulk of its input from the lateral geniculate nucleus of the thalamus, and is subdivided into blobs and interblobs for processing color and form, respectively [2.158]. Primary sensory areas like V1 and the primary somatosensory cortex (S1) have a topographic map of the body or visual field, where adjacent neurons represent adjacent points in space [2.159]. Higher-order association areas, like the prefrontal cortex, have a more abstract, distributed organization, integrating information from multiple modalities [2.160].

This cytoarchitectonic non-uniformity is tightly linked to functional specialization. The hierarchical organization of the cortex, from primary sensory areas to multimodal association areas, allows for a progressive transformation of information. A spot of light on the retina is transformed into a line, then a shape, then a recognized object through successive stages of processing in different cortical areas [2.161]. This flow of information is not strictly linear but involves massive recurrent connections, both within and between areas, allowing for feedback and dynamic adjustment of processing. The non-uniform cortex, therefore, is not a blank slate but a highly structured, pre-wired computer, where its physical layout encodes the algorithm for transforming raw sensory input into a rich, meaningful model of the world.

### 2.6.3 Asymmetric Left and Right Brain Hemispheres

The two cerebral hemispheres, while appearing roughly symmetrical, are fundamentally asymmetric in both structure and function, a phenomenon known as lateralization [2.162]. This asymmetry is one of the most dramatic examples of non-uniformity in the brain. The most well-known example is language dominance. In approximately 95% of right-handed individuals and 70% of left-handed individuals, the left hemisphere is dominant for language processing, housing critical areas like Broca's area (speech production) and Wernicke's area (language comprehension) [2.163]. Damage to the left hemisphere is far more likely to cause aphasia than equivalent damage to the right. The right hemisphere, in contrast, is generally dominant for visuospatial processing, facial recognition, and processing the prosody (emotional tone) of speech [2.164].

Structural correlates of this functional asymmetry exist. The planum temporale, a region within Wernicke's area, is typically larger in the left hemisphere [2.165]. The Sylvian fissure, which houses this region, is also longer and less angled on the left side. This anatomical difference is present from birth, suggesting a strong genetic basis for lateralization [2.166]. Beyond language, the hemispheres show biases in emotional processing, with the right hemisphere playing a more dominant role in expressing and recognizing emotions [2.167]. Motor control is also lateralized, with the left hemisphere controlling the right side of the body and vice versa, due to the decussation of fibers in the brainstem [2.168].

The evolutionary advantage of this asymmetry is thought to be increased processing efficiency. By specializing the hemispheres, the brain can avoid duplicating complex functions, freeing up neural resources for other tasks [2.169]. It allows for parallel processing of different types of information. However, this division of labor requires seamless communication. The corpus callosum, a bundle of over 200 million axons, facilitates this exchange, allowing the two hemispheres to integrate their specialized operations into a unified conscious experience [2.170]. Conditions like split-brain syndrome, where the corpus callosum is severed, dramatically demonstrate the consequences of this separation, with each hemisphere appearing to have its own independent perceptions and intentions [2.171]. This profound asymmetry underscores that the brain's non-ideal structure is not random but a highly evolved solution to the problem of managing complex cognitive functions.

**2.6.4 Diverse Cross-Regional Connections**

The functional coherence of the brain arises not from the isolated operation of its regions but from the diverse and specific connections that link them [2.172]. These cross-regional connections form a network with unparalleled complexity. The brain employs a variety of long-range projection neurons, whose axons traverse vast distances through the white matter. The nature of these connections is incredibly diverse. Some are unidirectional, carrying information from a lower to a higher region in a hierarchy, like the ascending sensory pathways [2.173]. Others are bidirectional, with separate populations of neurons sending signals forward (feedforward) and backward (feedback) [2.174]. Still others are part of closed-loop circuits, such as the cortico-striato-thalamo-cortical loops, which are involved in action selection and habit formation [2.175].

The diversity of these connections is also reflected in their neurotransmission. Most long-range projections are glutamatergic (excitatory), but there are also long-range inhibitory GABAergic projections, particularly from the cortex to subcortical structures [2.176]. The strength and plasticity of these connections are dynamically regulated. For example, the hippocampus is thought to rapidly encode new episodic memories, which are then slowly transferred and consolidated into the neocortex during sleep through a process that depends on coordinated replay and the strengthening of specific cross-regional connections [2.177]. This dialogue between the hippocampus and neocortex is a prime example of how diverse connectivity supports high-level cognitive functions.

Modern techniques like diffusion tensor imaging (DTI) and resting-state functional MRI (fMRI) have allowed scientists to map these connections on a macroscale, revealing the brain's connectome as a network with hubs of high connectivity (like the posterior cingulate cortex in the default mode network) and bridges between modules [2.178]. This diverse connectivity enables a range of dynamic phenomena, such as the flexible recruitment of brain regions for different tasks and the integration of information across modalities. The very possibility of having a "memory" or a "thought" that draws on information stored in disparate regions—visual, auditory, emotional—relies entirely on the existence of these diverse cross-regional highways. The non-ideal factors here is not a flaw but the essence of the system: a distributed, interconnected network that transcends the limitations of any single, uniform processor.

**2.6.5 Diverse Heterogeneous Left-Right Brain Hemisphere Connections**

The connection between the left and right hemispheres, primarily mediated by the corpus callosum but also by smaller commissures like the anterior and posterior commissures, is itself a site of immense diversity and heterogeneity [2.179]. Far from being a uniform cable, the corpus callosum carries millions of axons that are topographically organized. Fibers from specific cortical areas cross to their homologous counterparts on the opposite side. For instance, motor fibers from the hand area of the left motor cortex cross to the right motor cortex, and visual fibers from the left visual field (processed by the right hemisphere) cross to the left hemisphere [2.180].

However, this organization is not perfect or complete. Some cortical areas are more strongly connected than others, and there is evidence of functional asymmetry in the callosal connections themselves [2.181]. Moreover, the corpus callosum transmits different types of information. It relays motor coordination signals, allowing for bimanual tasks, integrates sensory information (e.g., combining touch from the left hand with vision from the right visual field), and shares cognitive content [2.182]. The transmission speed can vary, with thicker, myelinated axons conducting faster than thinner ones. This heterogeneity in the callosal pathway means that inter-hemispheric communication is not instantaneous or uniform; it can have a latency of up to 100 milliseconds [2.183].

This diverse and heterogeneous connection is critical for creating a unified mind from two physically separate hemispheres. It allows for the sharing of information necessary for coherent perception and action. For example, when reading, visual information from both sides of the page is integrated in both hemispheres via the corpus callosum [2.184]. The disruption of these connections, as in split-brain patients, reveals the independent processing streams of each hemisphere [2.185]. The diverse nature of these connections ensures that the integration is not a simple averaging but a sophisticated negotiation, where each hemisphere can specialize and contribute its unique strengths to the overall cognitive process, a fundamental non-ideal factors that underpins the brain's remarkable capacity for integrated thought and behavior.

## 2.7 Developmental and Environmental Origins of Neural Non-Ideal factors

## 2.7.1 Critical Period Mismatch and Its Lifelong Consequences

Critical periods are restricted developmental windows during which the brain exhibits heightened plasticity and is exquisitely sensitive to specific environmental inputs for shaping neural circuits [2.186]. The canonical example is ocular dominance plasticity in the primary visual cortex, where the balanced input from both eyes is necessary for normal binocular vision. If one eye is deprived of vision during this critical period, the cortical territory responsive to that eye shrinks, leading to amblyopia ("lazy eye") [2.187]. This mismatch, where the neural circuit fails to develop properly due to inadequate or inappropriate environmental input, has profound and often irreversible lifelong consequences.

The timing of critical periods is genetically programmed but can be influenced by experience. A mismatch occurs when the required environmental input is not available at the precise time the brain is prepared to receive it. For example, the acquisition of a native accent in a second language is most efficient in childhood and becomes markedly more difficult after puberty, suggesting a closing of a critical period for phonological learning [2.188]. This mismatch is not a personal failing but a consequence of the brain's biological schedule. Early adversity, such as neglect or abuse, can disrupt the normal unfolding of multiple critical periods for attachment, emotion regulation, and language development [2.189]. This can lead to a cascade of problems, including increased risk for psychiatric disorders, difficulties in social relationships, and cognitive deficits in adulthood. The brain's non-ideal factors here is its rigidity at certain times; its peak adaptability is a double-edged sword, as it leaves the system vulnerable to permanent malformation if the correct inputs are missing.

## 2.7.2 Perception-Dependent Divergence in Neural Circuit Formation

An individual's unique perceptual experiences during development are a primary driver of divergence in neural circuit formation, leading to the non-ideal, idiosyncratic wiring of the brain [2.190]. The brain is not born with a fully formed map of the world; it constructs this map through active engagement with its environment. This process, guided by principles of Hebbian plasticity, ensures that neural circuits are tailored to the individual's specific life history [2.191]. For instance, a child growing up in a visually rich environment with diverse textures and patterns will develop a visual cortex with a different statistical distribution of receptive fields compared to a child raised in a visually impoverished setting [2.192]. Similarly, musicians who begin training early show enlarged cortical representations for finger movements and enhanced auditory processing, a direct consequence of the intensive, specialized perceptual input they receive [2.193].

This divergence is not limited to the senses. The development of the self-concept and social cognition is heavily influenced by perceptual input from caregivers and peers. A child who perceives consistent, responsive care will develop secure attachment circuits, while a child who perceives inconsistent or frightening care will develop anxious or avoidant attachment patterns [2.194]. These early perceptual templates become ingrained in the neural architecture of the limbic system and prefrontal cortex, shaping emotional regulation and relationship styles for life. This principle extends to learning: a student who consistently perceives their teacher as supportive is more likely to develop a growth mindset and engage in self-regulated learning, while a student who perceives their teacher as judgmental may develop learned helplessness [2.195]. Thus, the brain's "non-ideal" wiring is, in large part, a reflection of its unique perceptual journey through the world.

### 2.7.3 Movement-Dependent Divergence in Neural Circuit Formation

Just as perception shapes the brain, so too does movement. The principle of "motor equivalence" suggests that the brain learns skills, not muscle commands [2.196]. This means that the neural circuits for a skill are refined through practice, but the specific motor program can vary. However, the overall pattern of movement during development has a profound impact on circuit formation. Physical play, sports, dance, and even fidgeting contribute to the development of the cerebellum, basal ganglia, and motor cortex [2.197]. Repetitive movements strengthen the relevant corticospinal and subcortical pathways, while the lack of diverse movement can lead to weaker or less refined circuits [2.198]. Children who are physically active tend to have better-developed proprioceptive systems, which inform the brain about body position, leading to more coordinated and graceful movement [2.199].

This movement-dependent divergence has long-term consequences. Research on embodied cognition suggests that physical metaphors are deeply embedded in thought; for example, the physical sensation of weight is linked to the abstract concept of importance [2.200]. A child with rich physical experiences may therefore develop a more grounded, intuitive understanding of abstract concepts. Furthermore, the development of fine motor skills, like handwriting or playing a musical instrument, is closely linked to the maturation of frontal lobe circuits involved in attention and planning [2.201]. The process of mastering a complex motor skill requires sustained attention, error correction, and iterative refinement—all hallmarks of self-regulated learning. Therefore, the motor system is not just for movement; it is a fundamental trainer for the cognitive control systems that govern learning and behavior across the lifespan.

### 2.7.4 Emotion-Dependent Divergence in Neural Circuit Formation

Emotional experiences are perhaps the most potent sculptors of neural circuits, especially during early development [2.202]. The brain's limbic system, particularly the amygdala and its connections to the prefrontal cortex and hippocampus, is highly sensitive to emotional valence [2.203]. Positive emotional experiences, such as warmth, safety, and joy, promote the growth of healthy neural connections, enhance synaptic plasticity, and buffer against stress [2.204]. These experiences are associated with optimal development of the orbitofrontal cortex, a region critical for emotion regulation and social decision-making [2.205].

Conversely, chronic negative emotions, such as fear, anxiety, and trauma, can have a deleterious impact. Prolonged exposure to stress hormones like cortisol can impair neurogenesis in the hippocampus, weaken the inhibitory control of the prefrontal cortex over the amygdala, and sensitize the HPA axis, creating a hyper-reactive stress response system [2.206]. This leads to a fundamental divergence in brain architecture: one brain may be optimized for exploration, trust, and learning, while another may be perpetually poised for threat detection and defensive reactions [2.207]. This emotional dependency explains why two individuals can have vastly different reactions to the same event. Their brains have been shaped by a lifetime of divergent emotional experiences, creating unique internal landscapes that interpret the world through the lens of their past feelings. This divergence is the neural substrate of personality and psychopathology.

### 2.7.5 Thought-Dependent Divergence in Neural Circuit Formation

The act of thinking itself is a powerful force in shaping the brain. This is the principle of neuroplasticity in action: "neurons that fire together, wire together" [2.208]. The thoughts an individual chooses to entertain, ruminate on, or master become physically encoded in the strength and configuration of their neural networks. A student who practices logical reasoning and critical analysis strengthens the dorsolateral prefrontal cortex [2.209]. A person who engages in daily meditation or mindfulness cultivates circuits for attentional control and meta-awareness, often reflected in increased gray matter density in the insula and anterior cingulate cortex [2.210]. The famous London taxi drivers study showed that the posterior hippocampus, involved in spatial navigation, is larger in drivers with more years of experience, a direct result of the intense navigational thinking they perform [2.211].

This thought-dependent divergence creates intellectual specializations. The neural circuitry of a mathematician, trained to think in terms of abstract symbols and formal proofs, will be distinctly different from that of a poet, trained to think in metaphors and emotional resonances [2.212]. Even within a field, the depth of thought on a specific topic will deepen and refine the relevant circuits. This is the biological basis of expertise. It underscores that the brain is not a passive vessel waiting to be filled but an active participant in its own construction. The quality and quantity of one's thoughts directly mold the physical structure of the organ responsible for thought, creating a recursive loop of self-directed brain development. The non-ideal nature of this process is its openness to individual choice and discipline.

### 2.7.6 Behavior-Dependent Divergence in Neural Circuit Formation

Behavior, the outward manifestation of internal neural processes, also feeds back to shape those very processes [2.213]. This behavior-dependent divergence is a core tenet of operant conditioning and self-regulated learning. When an individual repeatedly performs a specific behavior, the neural circuits underlying that behavior are reinforced through a process of long-term potentiation (LTP) [2.214]. This applies to both desirable habits, like studying diligently or exercising regularly, and undesirable ones, like procrastinating or smoking. Each repetition of the behavior strengthens the association between the contextual cues and the motor and cognitive routines, eventually making the behavior more automatic and less reliant on conscious effort [2.215].

This principle is evident in the development of skills. A programmer who spends hours writing code strengthens the neural pathways for problem decomposition, syntax recall, and debugging [2.216]. An athlete who practices a specific play improves the precision and timing of their motor commands [2.217]. The "deep work" capability praised in the workplace is not an innate talent but a behavioral habit that has been cultivated through consistent practice of focused, uninterrupted work [2.218]. This practice physically alters the brain, increasing the efficiency of attentional control systems. Conversely, habitual multitasking, such as constantly switching between tasks or checking notifications, trains the brain for distraction, weakening the prefrontal cortex's ability to maintain focus [2.219]. This creates a significant divergence in cognitive capacity, where one individual's brain is optimized for depth and another for breadth, but at the cost of shallow processing. The brain becomes a mirror of the behaviors it has most frequently enacted.

**2.7.7 Experience-Dependent Divergence in Neural Circuit Formation**

Finally, the overarching theme of all the previous subsections is experience-dependent plasticity [2.220]. Every moment of an individual's life—their perceptions, movements, emotions, thoughts, and behaviors—contributes to a unique constellation of neural activity that, over time, sculpts their brain. This is the ultimate origin of neural non-ideality. No two individuals have identical experiences, even siblings raised in the same household. Differences in birth order, parental treatment, friendships, chance encounters, and illnesses all contribute to a divergent path of neural development [2.221]. This divergence is not a deviation from a norm but the creation of a personalized norm. The brain is designed to be malleable, to absorb the lessons of its environment and adapt accordingly.

This process continues throughout life. While plasticity is highest in youth, the adult brain retains a remarkable capacity for change. Learning a new language, recovering from a stroke, or undergoing psychotherapy all involve experience-dependent rewiring of neural circuits [2.222]. This lifelong plasticity ensures that the brain is never a finished product. It's non-ideal, idiosyncratic structure is a dynamic record of a lifetime of living, a testament to the profound truth that we are, quite literally, what we have experienced.

**2.8 Evolutionary Rationale for Neuronal Inconsistency**

**2.8.1 Robustness through Functional Redundancy and Distributed Coding**

From an evolutionary perspective, the inconsistencies and non-idealities of the neural substrate are not accidents of biology but features that have been selected for their contribution to survival and adaptability [2.223]. A key rationale for neuronal inconsistency is the enhancement of system robustness. Biological systems, unlike engineered ones, must survive in a hazardous and unpredictable world. A neural system built from perfectly identical, highly optimized components would be elegant but brittle; the failure of a single component could lead to catastrophic system failure [2.224]. In contrast, the brain's functional redundancy and distributed coding provide a powerful defense against such fragility [2.225].

Redundancy is evident at multiple levels. Multiple sensory organs (two eyes, two ears) provide backup for each other. Within the nervous system, multiple neural pathways can mediate similar functions; for example, there are several descending motor pathways from the cortex to the spinal cord [2.226]. This ensures that if one pathway is damaged, others can partially compensate. At the level of information coding, distributed representations are far more robust than localist ones [2.227]. If a single neuron in a localist code dies, the entire concept it represents is lost. In a distributed code, where a memory or concept is represented by a pattern of activity across thousands of neurons, the loss of a few neurons degrades the signal slightly, like removing a few pixels from a photograph, but the overall information remains intact [2.228]. This fault tolerance is a direct benefit of heterogeneity, as a diverse population of neurons is less likely to fail in the same way. This principle of "graceful degradation" is a hallmark of biological intelligence and a key reason why the brain can continue to function after injury, a resilience that artificial systems still struggle to match [2.229].

### 2.8.2 Flexibility Enabled by Specialized Neuron Types

Evolution has favored flexibility, the ability to adapt to novel and changing environments [2.230]. The existence of numerous specialized neuron types is a primary engine of this flexibility [2.231]. Rather than relying on a single type of universal processor, the brain employs a vast array of specialized cells, each exquisitely tuned to a specific task. The diversity of interneurons in the cortex—parvalbumin-positive, somatostatin-positive, vasoactive intestinal peptide-positive, and many others—creates a rich palette of inhibitory control mechanisms [2.232]. These different interneuron types target different parts of the postsynaptic neuron (soma, axon initial segment, dendrites) and have different temporal dynamics (fast-spiking, adapting, irregular-spiking) [2.233]. This allows the network to generate a wide repertoire of oscillatory rhythms and to implement sophisticated forms of gain control and filtering [2.234].

This specialization allows neural circuits to be configured on the fly for different cognitive demands. A network can switch from a state of high excitability suitable for sensory detection to a state of high inhibition suitable for focused attention, simply by recruiting different populations of inhibitory interneurons [2.235]. This kind of operational flexibility would be impossible with a homogeneous population of neurons. It is analogous to an army equipped with infantry, cavalry, artillery, and special forces, able to deploy the right unit for the right mission, rather than an army of only one type of soldier [2.236]. The cost of maintaining this diversity is high in terms of energy and genetic complexity, but the payoff in terms of behavioral adaptability is immense, allowing organisms to solve a vast range of problems with a single, general-purpose organ [2.237].

### 2.8.3 Adaptability via Population-Level Signal Averaging and Noise Filtering

A third evolutionary rationale for neuronal inconsistency is adaptability, the ability to learn and improve over time [2.238]. The brain achieves this through mechanisms that exploit population-level statistics. The variability in individual neuron responses (neural noise) is not simply discarded but can be used as a resource [2.239]. By averaging the responses of a large population of neurons, the brain can extract a more reliable and accurate estimate of a stimulus than any single neuron could provide [2.240]. This is known as the "law of large numbers" in neural coding. The random fluctuations (noise) in each neuron's response are independent, so when averaged, they tend to cancel out, leaving a clearer "signal" [2.241].

This population averaging is also a powerful noise-filtering mechanism. The brain can distinguish a true, consistent signal from random fluctuations by looking for correlated activity across many neurons [2.242]. This makes the system robust to the failure of individual components and allows it to operate effectively in noisy environments. Furthermore, this population-level approach enables probabilistic inference [2.243]. The brain does not compute with certainty but with probabilities, constantly updating its beliefs based on incoming evidence. The distributed and noisy nature of neural representations is ideally suited for implementing Bayesian inference, where uncertainty is explicitly represented [2.244]. This allows for adaptive decision-making under uncertainty, a critical skill for survival. Thus, the brain's non-ideal, noisy components are not a liability but are harnessed by evolution to build a system that is not only robust and flexible but also inherently capable of learning and adapting to an uncertain world [2.245].

## 2.9 Technological Challenges in Modeling and Measuring Neuronal Heterogeneity

### 2.9.1 Limitations of Current Brain-Machine Interfaces in Resolving Single-Neuron Diversity

Brain-machine interfaces (BMIs) hold great promise for restoring function in neurological disorders and advancing our understanding of the brain [2.246]. However, a major technological hurdle is their inability to resolve the full diversity of single neurons [2.247]. Most clinical BMIs, such as those used for controlling prosthetic limbs or cursor movement, rely on electrodes that record the spiking activity of small populations of neurons, often referred to as "multi-unit activity" [2.248]. These devices typically sample from a few hundred neurons at best, and the recording sites are relatively coarse. They cannot reliably isolate and track the activity of individual neurons over long periods, let alone characterize their diverse morphologies, receptor profiles, or electrophysiological subtypes [2.249].

This limitation forces BMIs to use population-level signals for decoding intention. While effective for gross motor control, this approach ignores the rich heterogeneity of the neural code [2.250]. A BMI that could selectively read from or write to specific classes of neurons (e.g., only fast-spiking interneurons or only pyramidal cells in a specific layer) would be far more precise and efficient [2.251]. The inability to resolve single-neuron diversity means that BMIs are essentially operating blind to the most fundamental aspect of neural computation [2.252]. They treat the brain as a black box of indistinguishable units, limiting the sophistication and nuance of the interaction. Overcoming this challenge requires the development of higher-density electrode arrays, advanced signal processing algorithms for spike sorting, and perhaps even optical or molecular methods for interfacing with specific cell types [2.253].

### 2.9.2 Spatial Resolution Constraints in Recording Deep Brain Structures

Our understanding of the brain is heavily biased towards its superficial layers, particularly the cortex, due to the difficulty of accessing deep brain structures with high-resolution recording techniques [2.254]. Methods like functional MRI (fMRI) have excellent whole-brain coverage but poor spatial resolution, typically measuring voxels of several cubic millimeters, which contain millions of neurons and multiple cell types [2.255]. Techniques with higher spatial resolution, such as two-photon microscopy, are limited to superficial depths of a few hundred microns, primarily in the cortex of exposed brains in animal models [2.256].

This creates a significant blind spot for understanding the role of subcortical nuclei, such as the thalamus, basal ganglia, amygdala, and hypothalamus, which are buried deep within the brain and play critical roles in motor control, emotion, motivation, and homeostasis [2.257]. Our knowledge of their detailed neural dynamics and circuit interactions is severely limited. New technologies are emerging to address this, such as endoscopic microscopes and miniaturized fluorescence microscopes (miniscopes) that can be implanted to image deep structures [2.258]. However, these methods still face challenges in scalability, invasiveness, and long-term stability [2.259]. The inability to record from these deep structures with the same fidelity as the cortex means that our models of brain function are incomplete, lacking a critical piece of the puzzle. This spatial resolution constraint is a fundamental bottleneck in achieving a truly holistic understanding of the brain's non-ideal architecture [2.260].

### 2.9.3 Noise and Interference in Neural Signal Acquisition

Even when a signal is successfully acquired, it is invariably contaminated by various forms of noise and interference, adding another layer of technological non-ideal factors to the measurement process [2.261]. This noise can be biological or technical. Biological noise includes electromyographic (EMG) signals from muscle activity, electrocardiographic (ECG) artifacts from the heartbeat, and movement artifacts from the subject [2.262]. Technical noise includes electromagnetic interference from power lines (60 Hz hum), radiofrequency interference from wireless devices, and thermal noise within the recording equipment itself [2.263]. These contaminating signals can be orders of magnitude larger than the neural signals of interest, especially in non-invasive recordings like EEG [2.264].

Sophisticated signal processing techniques, such as filtering, artifact rejection, and independent component analysis (ICA), are essential to clean the data [2.265]. However, these methods are not perfect and can sometimes remove genuine neural signal along with the noise [2.266]. This creates a significant challenge for accurately interpreting the data. For instance, an observed "desynchronization" in an EEG signal could be a true cognitive process or simply an artifact of the subject clenching their jaw [2.267]. The presence of this noise and interference means that the data we collect is always an imperfect representation of the underlying neural reality, forcing us to make cautious interpretations and acknowledge the limitations of our tools [2.268].

**2.9.4 Computational Complexity in Decoding Heterogeneous Neural Signals**

The final and perhaps most daunting challenge is the computational complexity of decoding the brain's heterogeneous neural signals [2.269]. The brain's computational power arises from the massively parallel, nonlinear interactions of billions of neurons, each with its own complex dynamics [2.270]. Simulating even a fraction of this complexity is a Herculean task. A simulation of a single cortical column containing 100,000 neurons can require supercomputing resources [2.271]. Scaling this up to a whole brain with 86 billion neurons is currently beyond the reach of any existing technology and may remain so for decades [2.272].

Decoding the meaning of neural activity in real-time for applications like BMIs or brain-inspired computing is equally complex [2.273]. Algorithms must identify patterns in high-dimensional data streams, infer cognitive states, and predict intentions, all in the face of noise and variability [2.274]. This requires immense computational power and sophisticated machine learning models, such as deep neural networks [2.275]. Ironically, these AI models are inspired by the brain but require far more energy and computational resources to run than the biological brain itself [2.276]. The computational complexity is a fundamental barrier to creating realistic models of the brain that incorporate all its non-ideal factors, forcing researchers to use simplifying assumptions and approximations that inevitably lose important details of the biological reality [2.277].

**2.10 Implications for Artificial Intelligence and Neuromorphic Engineering**

**2.10.1 Challenges in Replicating Biological Heterogeneity in Silicon Systems**

Current artificial neural networks (ANNs), despite their success, are fundamentally different from their biological counterparts in their lack of heterogeneity [2.278]. ANNs are built from millions of identical, simplistic processing units (artificial neurons) arranged in regular layers [2.279]. This uniformity makes them easier to train and scale with current software and hardware. However, it deprives them of the robustness, flexibility, and adaptability conferred by biological neuronal inconsistency [2.280]. Replicating biological heterogeneity in silicon systems is a significant challenge [2.281]. It requires designing and fabricating chips with diverse types of artificial neurons, each with different activation functions, time constants, and connection rules [2.282]. This moves away from the current paradigm of homogeneous computing and into the realm of specialized, heterogeneous computing architectures [2.283].

Such designs would be far more complex to engineer, test, and manufacture. The yield rate for producing chips with intentional, functional variations would likely be lower [2.284]. Moreover, training algorithms for these heterogeneous ANNs would need to be rethought. Standard backpropagation assumes identical units; a new learning theory would be needed to manage the optimization of a system with diverse components [2.285].

Projects like IBM's TrueNorth and Intel's Loihi are early steps in this direction, incorporating elements of spiking communication and localized learning [2.286]. However, they still fall far short of the true diversity found in biology [2.287]. Embracing biological heterogeneity could lead to more robust and intelligent AI, but it requires a radical shift in how we design and think about computing systems [2.288].

## 2.10.2 Energy Efficiency Gaps between Biological and Artificial Networks

The most glaring discrepancy between biological and artificial networks is energy efficiency [2.289]. The human brain, weighing about three pounds, consumes approximately 20 watts of power—roughly the energy of a dim lightbulb [2.290]. In stark contrast, the largest AI models today can consume megawatts of power during training, requiring entire server farms cooled by massive air-conditioning systems [2.291]. This energy efficiency gap of many orders of magnitude is a critical issue for the sustainability and practical deployment of AI [2.292]. The brain's efficiency stems from several non-ideal features that are difficult to replicate in silicon [2.293]. First, it uses analog, event-driven processing. Neurons only expend energy when they fire, and communication is sparse, with only a small fraction of neurons active at any given time [2.294]. Digital computers, however, operate on a clock cycle, burning energy continuously, and typically use dense matrix multiplications, even when most values are zero [2.295].

Second, the brain's 3D architecture allows for extremely short wiring lengths, minimizing energy loss from signal transmission [2.296]. In contrast, 2D silicon chips suffer from the "interconnect bottleneck," where moving data between the CPU and memory consumes more energy than the computation itself [2.297]. Finally, the brain's use of mixed-signal processing (both analog and digital-like spikes) and its ability to function with significant noise and imprecision are key to its low power consumption [2.298]. Artificial systems strive for perfection and determinism, which is energy-intensive [2.299]. Future neuromorphic engineering aims to bridge this gap by building brain-inspired chips that use asynchronous, event-driven, and mixed-signal designs to achieve unprecedented energy efficiency, potentially unlocking new applications in mobile and embedded AI [2.300].

## 2.10.3 Structural and Dynamic Limitations of 2D vs. 3D Neural Architectures

The physical architecture of the brain—a soft, 3D gel of neurons and glia packed with nanometer-scale wires (axons and dendrites)—stands in stark contrast to the rigid, 2D plane of silicon chips [2.301]. This 3D structure is a fundamental non-ideal factors that provides significant advantages [2.302]. It allows for an ultra-high packing density and minimizes the distance between connected components. The brain's wiring is optimized for minimal conduction delay and energy cost, a principle known as "wiring economy" [2.303]. In contrast, 2D silicon architectures suffer from a fundamental limitation: the separation between logic (CPU) and memory (RAM). Data must constantly be shuttled back and forth over long distances, which is slow and energy-intensive. This is the von Neumann bottleneck [2.304].

While stacked 3D chips and advanced packaging techniques are being developed, they still cannot match the true 3D integration of the brain [2.305]. The brain's structure is also dynamic; it can grow new connections and prune old ones, something a static silicon chip cannot do [2.306]. The dynamic nature of the brain's architecture allows it to be reconfigured on the fly for different tasks [2.307]. This structural and dynamic flexibility is a key component of its adaptability. Artificial systems are constrained by their physical form, limiting their ability to evolve and self-repair [2.308]. Overcoming this limitation may require a paradigm shift towards bio-integrated electronics or soft robotics, where the boundary between the machine and its environment becomes more fluid [2.309]. Until then, the 3D, dynamic, and self-assembling nature of the biological brain will remain a significant source of inspiration and a formidable challenge for engineers [2.310].

## 2.11 Conclusion: Embracing the "Messy" Brain as an Adaptive System

### 2.11.1 Neuronal Inconsistency as a Core Feature, Not a Flaw

In conclusion, the myriad non-ideal factors cataloged in this chapter—from the morphological quirks of individual neurons to the systemic asymmetries of the whole brain—are not defects to be lamented but the very foundation of biological intelligence [2.311]. The traditional scientific impulse has been to seek an idealized, noise-free, perfectly optimized model of the brain [2.312]. This chapter argues for a radical paradigm shift: to embrace the "messy" brain as an adaptive system where inconsistency is a core feature, not a flaw [2.313]. The diversity, noise, and idiosyncratic connectivity that define neural systems are the products of millions of years of evolution, carefully tuned to create a machine that is robust, flexible, and adaptable in a chaotic world [2.314]. This perspective reframes our understanding of neural computation, suggesting that intelligence emerges not from the elimination of noise but from its sophisticated management and exploitation [2.315].

### 2.11.2 Integration of Multiple Non-Ideal Factors in Realistic Brain Models

Moving forward, the challenge for neuroscience is to integrate these multiple non-ideal factors into a new generation of realistic brain models [2.316]. This means abandoning overly simplified abstractions and confronting the full complexity of the biological system [2.317]. Future models must account for the interplay between molecular noise, cellular diversity, and large-scale network architecture [2.318]. This will require interdisciplinary collaboration, bringing together experimental neuroscientists, theorists, and engineers [2.319]. The ultimate goal is not just to simulate the brain but to understand the principles of intelligence that emerge from its non-ideal structure [2.320]. By learning from the brain's ingenious solutions to problems of robustness, efficiency, and adaptability, we can not only gain deeper insight into ourselves but also revolutionize the fields of artificial intelligence, medicine, and computing, paving the way for a new era of truly intelligent machines [2.321].

## 2.12 Expectations of Mathematical Analysis

Traditional mathematical models (e.g., linear differential equations, static network models) indeed struggle to capture the nonlinear dynamics(e.g., chaotic firing), large-scale parallelism(86 billion neurons × $10^3$ synapses), and energy efficiency constraints(occupying 2% of body weight yet consuming 20% of energy) of the biological brain, thus having descriptive limitations. However, in recent years, novel computational models inspired by neurosciencehave made significant progress and can partially simulate brain mechanisms:

### 2.12.1 Breakthroughs in Deep Learning (DL)

Convolutional Neural Networks (CNNs) mimic the hierarchical processing of the visual cortex [2.322] and approach human-level performance in image recognition tasks [2.323]. Transformer models draw on the attentional mechanisms of the prefrontal cortex for long-range dependency processing [2.324]. DeepMind's AlphaFold predicts protein structures via self-supervised learning, with its algorithm inspired by the spatial navigation coding of the hippocampus [2.325]- [2.326].

### 2.12.2 Biological Plausibility of Spiking Neural Networks (SNNs)

SNNs use spike timing as the information carrier, making them closer to the "event-driven" characteristics of biological neurons. For example, Eliasmith's team's Spaun model [2.327] integrates 2.5 million neurons to achieve coordination of visual recognition, memory storage, and working memory. Indiveri's neuromorphic chips [2.328] simulate ion channel dynamics with energy consumption only 1/1000 that of traditional computers. Recent advancements include Rueckauer et al.'s [2.329] demonstration of SNN-based Bayesian inference and Roy et al.'s [2.330] neuromorphic system for real-time sensory processing.

### 2.12.3 Limitations and Complementarity

These models still simplify key features of the biological brain (e.g., molecular regulation of synaptic plasticity, metabolic support from glial cells) and lack descriptions of higher-order functions like the emergence of consciousness 2.331]. Thus, "cannot fully describe" is more accurate than "cannot describe"—traditional models serve as the foundation, while novel models are gradually approaching biological realism [2.332].

# 3 Mathematical Challenges in neural Analysis and modeling

## 3.1 Noise and Error Coupling

The human brain, with its approximately 86 billion neurons and trillions of synaptic connections, represents the most intricate known system in the universe. A fundamental paradox underpins its operation: this pinnacle of biological computation functions not through the pristine, error-free signaling of a digital computer, but by harnessing an environment saturated with noise [3.1]. This inherent stochasticity, far from being a flaw to be eliminated, is deeply woven into the fabric of neural computation, posing profound mathematical challenges for analysis. Neural noise manifests at every scale, from the quantum jitter of ion channels to the variability in neurotransmitter release and fluctuations in membrane potential [3.2]. For instance, even when presented with an identical stimulus, the precise timing and pattern of action potentials (spikes) from a neuron can vary dramatically across trials [3.3]. This variability is not merely random error; it is an intrinsic property of the biological hardware. Modeling such systems requires moving beyond deterministic equations to embrace probabilistic frameworks like stochastic differential equations or Markov processes [3.4]. However, this transition exponentially increases complexity. The central challenge lies in disentangling signal from noise—determining which variations are meaningful information encoding different stimuli or internal states, and which are mere epiphenomena of biological messiness [3.5]. An analysis that fails to account for noise coupling may misinterpret dynamic range limitations as pathological conditions or overlook critical computational roles of variability.

This challenge is further compounded by the non-linear nature of neural dynamics. A small amount of noise can trigger significant shifts in network behavior due to bifurcation points, where tiny perturbations lead to qualitatively different stable states [3.6]. Such phenomena make predictive modeling exceptionally difficult, as errors propagate non-additively. Furthermore, what appears as "noise" at one level of observation might constitute the signal at another [3.7]. For example, the seemingly erratic firing patterns of individual neurons could reflect a population code where the collective activity carries information, rendering the variability of single units functionally irrelevant [3.8]. Therefore, a successful mathematical analysis must define its operational level carefully. Attempts to model whole-brain dynamics using high-fidelity noisy components result in computationally intractable models, while overly simplified models risk losing essential biological truths about how noise shapes perception, decision-making, and learning [3.9]. The pervasive presence of noise thus forces analysts to grapple with a trade-off between realism and tractability, making the development of robust analytical tools for noisy neural systems one of the foremost frontiers in computational neuroscience.

### 3.2 Neuronal Individuality, Heterogeneity, and Mismatch Introduce Analytical Complexity

A foundational assumption in many early neural network models was homogeneity: that neurons within a given class were essentially identical functional units [3.10]. Modern neuroscientific research has decisively overturned this notion, revealing a staggering degree of neuronal individuality and heterogeneity [3.11]. This diversity exists across multiple dimensions: morphological (in dendritic arborization, soma size), electrophysiological (in firing rates, thresholds, and adaptation properties), molecular (in receptor subtypes, ion channel expression), and connectivity (in pre- and post-synaptic partners) [3.12]. This heterogeneity is not randomness; it is a feature sculpted by evolution and experience to enhance computational capacity [3.13]. For instance, inhibitory interneurons exhibit remarkable specialization, with distinct subclasses like parvalbumin-positive, somatostatin-positive, and vasoactive intestinal peptide-positive cells, each playing unique roles in shaping cortical dynamics and enabling complex cognitive functions [3.14]. The failure to account for this biological reality introduces immense analytical complexity into any mathematical model.

Modeling such diverse populations necessitates abandoning simple mean-field approximations that treat large groups of neurons as uniform ensembles [3.15]. Instead, researchers must develop heterogeneous population models, vastly increasing the number of parameters and state variables [3.16]. This shift moves the field toward high-dimensional dynamical systems, where traditional analytical methods often falter. Moreover, the mismatch between neurons—where two cells intended to perform similar functions exhibit divergent properties—can lead to unexpected emergent behaviors in networks [3.17]. For example, slight differences in axonal conduction velocity or synaptic delay between parallel pathways can generate oscillatory dynamics or phase shifts that are critical for information processing but are absent in idealized models [3.18]. These biological imperfections become integral to function. Consequently, an accurate analysis must incorporate these variations explicitly, requiring sophisticated statistical descriptions and simulations capable of handling large-scale heterogeneity [3.19]. Failure to do so results in models that are mathematically elegant but biologically naive, unable to replicate the robustness and flexibility observed in real brains.

### 3.3 Impact of Diversity on Modeling and Interpretation

The profound impact of neural diversity extends beyond the construction of models to fundamentally shape their interpretation [3.20]. Models based on homogeneous assumptions tend to produce clean, interpretable outputs with clear cause-and-effect relationships [3.21]. In contrast, models incorporating realistic heterogeneity yield messy, variable results that better mirror empirical data but are significantly harder to interpret [3.22]. The relationship between a model parameter and its behavioral output becomes blurred by the interaction of thousands of other variables representing individual differences [3.23]. This makes reverse engineering brain function from data exceptionally challenging. When analyzing experimental recordings, a scientist cannot assume that all neurons of a particular type will respond identically to a stimulus [3.24]. Variability in response profiles could stem from underlying biological diversity rather than measurement error or changing network context.

This has direct implications for fields like neuromorphic engineering and artificial intelligence [3.25]. AI systems inspired by the brain have traditionally employed uniform artificial neurons, achieving impressive performance but lacking the fault tolerance and energy efficiency of their biological counterparts [3.26]. The realization that heterogeneity confers robustness suggests that future generations of AI should incorporate more diverse computational elements [3.27]. However, doing so complicates the design and debugging process, mirroring the challenges faced in neuroscience [3.28]. Interpreting the activity of a diverse artificial neural network is as difficult as interpreting the brain itself. Thus, embracing diversity forces a paradigm shift from seeking simple, universal principles to understanding principles of distributed, redundant, and adaptive computation [3.29]. It demands new metrics and visualization techniques to extract meaning from complex, high-dimensional datasets generated by both real and simulated neural systems.

### 3.4 Unpredictability of Connectivity

At the heart of neural analysis lies the connectome—the comprehensive map of neural connections [3.30]. Despite advances in electron microscopy and tracing techniques, predicting the precise wiring diagram of even a small brain region remains largely impossible [3.31]. Connectivity is shaped by a complex interplay of non-uniform 3D genetic programs, developmental cues, and lifelong experience-dependent plasticity [3.32]. While general anatomical rules exist (e.g., certain areas are reciprocally connected), the specific synapses formed between individual neurons appear highly idiosyncratic [3.33]. This unpredictability stems from the combinatorial explosion of possible connections: with billions of neurons, each forming thousands of synapses, the potential connectomes are astronomically numerous [3.34]. This lack of predictability renders purely deductive modeling futile. Analysts cannot start with first principles to derive the exact structure of a network.

Instead, research relies heavily on empirical data collection, which is slow, expensive, and often limited to small volumes of tissue [3.35]. Even when data is available, sampling bias is a major concern [3.36]. Mathematical analyses must therefore work with incomplete and uncertain knowledge of the underlying architecture [3.37]. This forces reliance on generative models that create plausible network topologies based on statistical rules derived from partial data [3.38]. However, validating these models against ground truth is extraordinarily difficult [3.39]. The unpredictable nature of connectivity means that seemingly minor differences in wiring can lead to drastically different network dynamics and functional outcomes, a phenomenon known as sensitivity to initial conditions [3.40]. This unpredictability is a core reason why replicating findings across individuals or species is so challenging and underscores the importance of studying neural circuits as emergent properties of complex systems rather than predetermined blueprints.

## 3.5 Excessive System Complexity

The sheer scale of the nervous system creates excessive system complexity that defies complete analytical description [3.41]. The number of interacting components—neurons, glia, synapses, signaling molecules—is so vast that simulating a human brain at full resolution is currently far beyond the capabilities of any supercomputer [3.42]. This complexity arises not just from the number of parts but from the non-linear, recursive, and multi-scale interactions between them [3.43]. Processes unfold simultaneously across temporal scales, from millisecond spike events to long-term structural changes over years, and spatial scales, from nanometer-sized receptors to meter-long axons [3.44]. Integrating these disparate levels into a unified mathematical framework is a monumental task.

Attempts to simplify by reducing dimensionality inevitably sacrifice important details [3.45]. For example, coarse-graining a network by grouping neurons into populations loses information about local microcircuits [3.46]. Conversely, focusing on detailed microcircuitry makes it impossible to see global brain states [3.47]. This complexity leads to the emergence of properties like chaos and self-organization, which are notoriously difficult to analyze and control [3.48]. Because of this, much of neural analysis operates in the realm of effective theories—models that capture salient features of the system without claiming to represent every detail [3.49]. However, determining which features are salient requires careful experimentation and insight, highlighting the tight coupling between theory and experiment in this field [3.50]. Ultimately, the excessive complexity means that a complete mathematical understanding of the brain may remain forever out of reach, forcing scientists to focus on specific subsystems or functions.

## 3.6 Interference and Distortion in Neural Processes

Neural signals do not travel through isolated channels; they propagate through a dense, interconnected network where interference and distortion are inevitable [3.51]. Electrical fields generated by one group of active neurons can influence the excitability of neighboring neurons—a phenomenon known as ephaptic coupling [3.52]. Similarly, neurotransmitters released into the synaptic cleft can diffuse and bind to receptors on nearby, unintended targets, causing spillover effects [3.53]. These forms of interference corrupt the fidelity of information transmission. From a mathematical standpoint, modeling such crosstalk requires incorporating additional terms for lateral inhibition, volume transmission, and electromagnetic interactions, which rapidly increase model complexity and are rarely included in standard neural network models [3.54].

### 3.6.1 Interference Impact on Analytical Relationships

Interference disrupts the assumed independence of neural signals, a common assumption in many analytical techniques like correlation analysis [3.55]. If the activity of neuron A influences neuron B not through a direct synaptic connection but via electrical field effects, an analysis might incorrectly infer a direct functional link [3.56]. This leads to erroneous conclusions about network topology and causal relationships [3.57]. Techniques designed to detect functional connectivity, such as Granger causality, can be misled by such indirect interactions [3.58]. To mitigate this, advanced statistical methods like partial coherence or conditional Granger causality are required to isolate direct from indirect effects [3.59]. However, these methods demand even more data and computational power, and their accuracy depends on correctly specifying the model order [3.60]. The presence of interference thus necessitates more sophisticated analytical approaches and a healthy skepticism towards simple interpretations of correlated activity.

### 3.6.2 Distortion in Neural Signal Transmission

Signal distortion occurs throughout the neural pathway [3.61]. Synaptic transmission is inherently unreliable, with a release probability typically less than one, leading to stochastic failures [3.62]. Axons can act as low-pass filters, attenuating high-frequency components of the signal [3.63]. Dendrites actively process incoming inputs through voltage-gated ion channels, transforming the signal before it reaches the soma [3.64]. This means the information encoded in the spike train at the axon hillock is not a faithful copy of the original sensory input [3.65]. Modeling this distortion requires detailed biophysical models of individual neurons, such as multi-compartmental cable models [3.66]. While powerful, these models are computationally intensive and impractical for large-scale simulations [3.67]. As a result, simpler point-neuron models dominate, accepting the limitation that they distort the very processes they aim to describe [3.68]. This gap between biological reality and mathematical abstraction represents a significant barrier to understanding how information is preserved or transformed during neural processing.

### 3.7 Complex Changes Over Time, Age, and Environment

Neural systems are not static; they undergo continuous change over multiple timescales [3.69]. On short timescales, synaptic strength fluctuates due to short-term plasticity mechanisms like facilitation and depression [3.70]. On longer timescales, structural plasticity allows for the growth and retraction of dendritic spines and axons [3.71]. Across the lifespan, the brain matures through well-defined developmental stages involving massive synaptic pruning and myelination, then ages with associated cognitive decline [3.72]. Furthermore, the environment continuously shapes the brain through learning and experience [3.73]. This dynamic nature presents a severe challenge for mathematical analysis, which often seeks stable, time-invariant solutions [3.74].

Models built from data collected at a single time point provide only a snapshot of a perpetually changing system [3.75]. Longitudinal studies are required to capture these dynamics, but they are logistically challenging and expensive [3.76]. Mathematically, incorporating time-varying parameters transforms ordinary differential equations into non-autonomous systems, which are much harder to analyze [3.77]. Capturing lifelong changes requires integrating processes across vastly different timescales, from milliseconds to decades, within a single coherent framework [3.78]. Additionally, environmental influences introduce uncontrolled variables that can confound analyses [3.79]. A model trained on data from one environment may fail to generalize to another, reflecting the brain's remarkable adaptability [3.80]. Therefore, any robust analytical approach must embrace the principle of plasticity, treating the brain not as a fixed machine but as a dynamic, evolving system whose past experiences are embedded in its present structure.

## 3.8 Other extensive Modeling Difficulties

Beyond the challenges already discussed, numerous other difficulties plague neural modeling [3.81]. One major issue is the problem of parameter estimation [3.82]. Biophysical models contain hundreds of parameters (conductances, capacitances, and reversal potentials) that are difficult to measure directly and can vary significantly between cells [3.83]. Fitting these models to data is an ill-posed inverse problem, often resulting in multiple parameter sets that produce similar outputs, a phenomenon known as non-identifiability [3.84]. Another difficulty is the curse of dimensionality: as models grow in complexity, the volume of data needed to adequately sample the state space grows exponentially, quickly becoming unattainable [3.85]. Validation of models against experimental data is also fraught with problems, as no single experiment captures all relevant aspects of brain function [3.86]. Finally, there is a persistent gap between levels of explanation—linking molecular events to cellular dynamics, to network behavior, and finally to cognition and behavior [3.87]. Bridging these gaps requires integrative multiscale models, which combine incompatible formalisms and face immense computational hurdles [3.88]. Collectively, these extensive difficulties underscore the fact that mathematical analysis of the brain remains an immature science, driven more by ingenuity and approximation than by rigorous, predictive theory [3.89].

# 4 Learning Principles without Mathematical Analytical Relations

## 4.1 Unavailability of Mathematical Models

The foundation of classical artificial intelligence and machine learning lies in the formulation of explicit mathematical models, where relationships between variables are defined by precise equations and functions [4.1]. This paradigm assumes that a system's behavior can be accurately predicted or optimized through analytical expressions derived from first principles or learned from data via gradient descent [4.2]. However, the biological brain operates under fundamentally different constraints, rendering such traditional mathematical modeling largely inapplicable to understanding its core learning mechanisms. The unavailability of comprehensive mathematical models for neural learning arises not from a lack of effort but from intrinsic properties of the nervous system: its staggering complexity, nonlinearity, and reliance on emergent dynamics rather than centralized computation [4.3].

Neurobiological evidence indicates that the human brain contains approximately 86 billion neurons, each forming thousands of synaptic connections, culminating in an estimated $10^{15}$ synapses [4.4]. This scale alone presents a combinatorial explosion far beyond the capacity of any existing mathematical formalism to describe with precision [4.5]. Unlike engineered systems designed for transparency and predictability, the brain evolved through natural selection to prioritize functional robustness over theoretical elegance [4.6]. Consequently, its operations cannot be reduced to a set of closed-form equations governing global network states. For instance, while simplified models like the Wilson-Cowan equations offer insights into population-level dynamics such as attractor states or oscillatory behavior [4.7], they fail to capture the microscale heterogeneity observed across individual neurons and synapses [4.8].

Recent experimental findings further underscore this limitation. A groundbreaking study revealed that within a single neuron, distinct dendritic branches follow different plasticity rules during learning tasks: some synapses adhered to Hebbian principles, while others exhibited changes independent of neuronal firing patterns [4.9]. This multi-rule architecture defies conventional mathematical frameworks, which typically assume uniform update Principles across all parameters [4.10]. Moreover, the dynamic nature of these interactions introduces time-varying dependencies that resist static modeling approaches [4.11]. Even advanced techniques such as bifurcation analysis and stochastic noise simulation provide only partial descriptions of local stability rather than complete predictive models [4.12].

Another critical factor is the role of molecular-scale processes in shaping synaptic efficacy. Mechanisms such as neurotransmitter release probability, receptor trafficking, and intracellular signaling cascades operate stochastically and are influenced by factors ranging from metabolic state to glial support [4.13]. These biochemical fluctuations introduce significant variability even among seemingly identical neurons, making deterministic predictions impossible [4.14]. Furthermore, developmental and environmental influences create unique "neural fingerprints" across individuals, ensuring no two brains implement learning identically [4.15]. As demonstrated by studies on perceptual-dependent circuit formation, early sensory experiences permanently alter connectivity patterns in ways that cannot be captured by generic equations [4.16].

Given these challenges, researchers have increasingly shifted focus toward qualitative and computational models that emphasize functional outcomes over analytical tractability [4.17]. Approaches rooted in nonlinear dynamical systems theory allow scientists to characterize broad classes of behavior—such as bistability, hysteresis, or chaos—without specifying exact parameter values [4.18]. Similarly, agent-based simulations enable exploration of collective phenomena emerging from simple interaction rules [4.19]. While these methods do not yield symbolic solutions, they facilitate hypothesis generation and experimental design in domains where formal mathematics falls short [4.20]. Thus, the absence of mathematical models does not imply ignorance; instead, it reflects a mature recognition that some systems must be understood through alternative epistemological frameworks grounded in observation, perturbation, and statistical inference [4.21].

## 4.2 Absence of Dynamic Logic and Mathematical Relationships

In engineered information processing systems, logical operations and state transitions are governed by well-defined rules encoded in hardware or software [4.22]. Boolean logic gates, finite-state machines, and algorithmic sequences ensure predictable progression from input to output based on explicit conditional statements [4.23]. In contrast, the brain lacks a centralized instruction set or universal syntax dictating how neural representations evolve over time [4.24]. Instead, its "dynamic logic"—if one may use the term—is distributed, probabilistic, and context-sensitive, arising spontaneously from the interplay of millions of concurrently active elements [4.25]. This absence of formal mathematical relationships means that neural computation cannot be described by discrete transition tables or truth functions, nor can it rely on error-correcting codes or checksums to maintain fidelity [4.26].

Evidence for this decentralized architecture comes from both anatomical and functional studies. Anatomically, cortical circuits exhibit massive convergence and divergence, with each neuron receiving inputs from diverse sources and projecting to multiple targets [4.27]. This architecture supports parallel processing and redundancy but undermines any attempt to map specific inputs directly onto outputs through linear causality [4.28]. Functionally, experiments using calcium imaging reveal that individual neurons participate in multiple overlapping ensembles depending on task demands [4.29]. For example, GABAergic interneurons in the mouse dorsomedial prefrontal cortex show higher inter-individual correlation compared to glutamatergic neurons, suggesting specialized roles in social information routing without fixed logical assignments [4.30].

Moreover, the temporal structure of neural communication departs radically from digital computing paradigms. Rather than relying on clock-synchronized updates, neural events unfold at various timescales—from sub-millisecond spikes to slow modulatory waves lasting seconds [4.31]. Spike-timing-dependent plasticity (STDP), a key mechanism underlying associative learning, depends critically on the relative timing of action potentials, yet the resulting weight changes are inherently noisy and subject to metaplastic regulation [4.32]. There is no equivalent to binary addition or multiplication; instead, integration occurs through graded membrane potentials, nonlinear summation at dendrites, and neuromodulator-gated plasticity windows [4.33]. These processes resist translation into standard arithmetic operations because their outcomes depend on history, location, and concurrent network activity [4.34].

The implications extend beyond basic computations to high-level cognitive functions such as decision-making and reasoning. Human judgments often violate axiomatic rationality, exhibiting biases such as loss aversion, anchoring effects, and framing dependence—phenomena difficult to reconcile with normative models of logic [4.35]. Neuroeconomic studies demonstrate that choices emerge from competition between valuation systems (e.g., ventral striatum) and control networks (e.g., lateral prefrontal cortex), rather than being computed according to utility maximization formulas [4.36]. Even abstract thought involves pattern completion and analogical transfer rather than formal deduction, suggesting that cognition operates more like associative memory retrieval than theorem proving [4.37].

This absence of formal logic does not imply randomness or inefficiency. On the contrary, the brain achieves remarkable performance in uncertain environments precisely because it leverages statistical regularities rather than rigid algorithms [4.38]. By exploiting correlations across modalities and timescales, it constructs probabilistic internal models capable of prediction and adaptation [4.39]. Machine learning research has begun to recognize similar advantages in deep neural networks trained end-to-end, where intermediate layers develop representations not explicitly programmed but discovered through exposure to data distributions [4.40]. Nevertheless, unlike artificial networks whose weights are adjusted via backpropagation—a process dependent on full gradient information—the brain must navigate without access to such global derivatives [4.41].

Ultimately, the brain's dynamic logic appears better characterized as a self-organizing complex system operating near criticality than as a programmable Turing machine [4.42]. Its operations resemble those of a fluid medium responding to boundary conditions rather than a circuit executing instructions [4.43]. Understanding this requires shifting perspective from syntax to semantics, from discrete steps to continuous flows, and from certainty to uncertainty quantification [4.44]. Researchers now employ tools from statistical physics, information theory, and causal inference to probe these emergent properties, recognizing that while mathematical relationships may be absent, statistical dependencies abound [4.45].

## 4.3 Impossibility of Mathematical Computation

Despite decades of progress in computational neuroscience, the idea that the brain performs mathematical computation in the sense understood by mathematicians and computer scientists remains untenable [4.46]. Traditional computation relies on precise numerical representations, sequential execution of operations, and guaranteed convergence to correct answers given sufficient resources [4.47]. Biological neurons, however, operate under severe physical and energetic constraints that preclude such deterministic processing [4.48]. Action potentials are inherently noisy, synaptic transmission fails intermittently, and ion channels fluctuate stochastically—all contributing to variability that would render conventional algorithms unreliable if implemented directly [4.49]. Instead of performing calculations, the brain approximates solutions through analog, population-coded dynamics that exploit parallelism and redundancy [4.50].

One illustrative example is the challenge of real-time signal processing. Consider the task of integrating visual motion cues to estimate heading direction during navigation. Classical approaches involve matrix transformations, coordinate frame conversions, and recursive filtering—all mathematically intensive procedures requiring floating-point precision [4.51]. Yet insects accomplish similarly sophisticated navigation with miniature brains consuming mere microwatts of power [4.52]. Research suggests they achieve this not through explicit calculus but via tuned neural responses shaped by evolution [4.53]. Direction-selective cells in the optic flow pathway respond preferentially to particular velocity vectors, effectively implementing a form of template matching without ever representing speed as a number [4.54].

Similarly, numerical cognition in humans shows little evidence of direct arithmetic implementation.

Functional MRI studies indicate that mental calculation engages parietal regions involved in spatial attention and magnitude estimation, suggesting that we manipulate quantities through analog mental number lines rather than digit-by-digit computation [4.55]. Behavioral data confirm this: reaction times increase logarithmically with operand size, consistent with searching along a continuous representation rather than executing discrete steps [4.56]. Errors also follow systematic patterns—such as the "problem size effect" and "tie effect"—indicative of associative retrieval from memory rather than rule-based derivation [4.57].

Furthermore, energy limitations impose fundamental constraints on computational feasibility. Each action potential consumes approximately $10^9$ ATP molecules, and the brain uses about 20% of the body's total energy despite accounting for only 2% of its mass [4.58]. This extreme metabolic cost necessitates highly efficient coding strategies [4.59]. Artificial neural networks running on conventional silicon consume orders of magnitude more energy per operation than their biological counterparts [4.60]. Neuromorphic engineering efforts inspired by neurodynamics aim to bridge this gap by emulating analog, event-driven computation, highlighting that efficiency stems from avoiding general-purpose digital computation altogether [4.61].

Even tasks that appear mathematical at the behavioral level may not involve internal computation. For instance, Bayesian inference—a cornerstone of modern theories of perception and decision-making—is often cited as evidence of probabilistic reasoning in the brain [4.62]. However, recent work argues that observed behaviors could result from heuristic shortcuts or implicit learning of environmental statistics without constructing formal likelihood functions or posterior distributions [4.63]. Neural implementations likely approximate Bayes-optimal decisions through adaptive gain modulation and recurrent network dynamics rather than manipulating probability densities symbolically [4.64].

Thus, the impossibility of mathematical computation in the literal sense underscores a deeper principle: the brain solves problems through embodied, situated engagement with the world rather than abstract symbol manipulation [4.65]. It learns to respond appropriately to stimulus patterns through experience-driven plasticity, producing behavior that is functionally adequate rather than formally correct [4.66]. This perspective shifts emphasis from replicating human-level AI via improved algorithms to reverse-engineering the brain's unconventional computational principles, potentially leading to new paradigms of intelligent systems [4.67].

## 4.4 Absence of Gradient Information

Backpropagation of error, the cornerstone of deep learning, depends crucially on the availability of gradients—partial derivatives indicating how small changes in each parameter affect the overall objective function [4.68]. This global feedback mechanism allows layered networks to assign credit or blame across arbitrary depths, enabling end-to-end optimization [4.69]. However, there is no known biological correlate for such a system [4.70]. Neurons receive limited local information and lack direct access to downstream error signals required for weight updates. The absence of gradient information poses a central paradox in neuroscience: How can a massively interconnected network learn hierarchical representations without a teacher providing detailed correction vectors?

Several arguments highlight the implausibility of biologically plausible backpropagation. First, the algorithm requires symmetric forward and backward pathways—a condition not met in mammalian anatomy [4.71]. Cortical pyramidal cells project predominantly in one direction, and retrograde signaling occurs through slower, diffuse chemical means unsuitable for rapid, precise gradient transmission [4.72]. Second, error signals would need to be calculated and routed simultaneously to every synapse in the network, demanding unrealistic coordination and bandwidth [4.73]. Third, the method violates Dale's principle, which posits that a neuron releases the same neurotransmitter at all its synapses, whereas backpropagation requires separate excitatory and inhibitory pathways for positive and negative gradient components [4.74].

**4.5 Global-Objective Backpropagation Learning Is Not Feasible**

The concept of optimizing a global objective function through backpropagation assumes a centralized cost metric and a coordinated update mechanism spanning the entire network [4.75]. In artificial neural networks, this manifests as minimizing a scalar loss value averaged over many training examples, with gradients flowing uniformly across layers. Biologically, such a scheme is profoundly implausible due to anatomical, physiological, and evolutionary constraints. The brain lacks a unified cost function or a master controller capable of broadcasting loss gradients throughout billions of neurons [4.76]. Instead, it must rely on decentralized, modular learning governed by local interactions and sparse supervisory signals.

Anatomically, the brain exhibits a highly distributed architecture with specialized modules handling distinct functions—visual processing, motor control, emotion regulation—each embedded within partially segregated hierarchies [4.77]. There is no central processor overseeing all cognitive operations; thus, no single entity computes a global loss. Even executive functions mediated by the prefrontal cortex operate reactively rather than proactively, integrating inputs from various subsystems to guide behavior without maintaining a comprehensive error landscape [4.78].

Physiologically, the mechanisms for transmitting information are ill-suited for synchronous gradient updates. Chemical synapses introduce delays of several milliseconds, and spike propagation speeds vary widely across fiber types [4.79]. Such temporal dispersion prevents the tight coordination required for simultaneous parameter adjustments. Moreover, synaptic plasticity itself is gated by numerous factors—including neuromodulators, astrocytic signals, and local metabolic conditions—that differ across brain regions and change dynamically with behavioral state [4.80].

Evolutionarily, organisms did not evolve to minimize abstract loss functions but to survive and reproduce in variable environments [4.81]. Natural selection favors heuristics that produce good-enough solutions quickly, rather than globally optimal policies derived through exhaustive search [4.82]. A predator does not calculate the shortest escape trajectory using gradient descent; it responds instinctively based on conditioned reflexes and innate biases [4.83]. This pragmatic orientation implies that learning proceeds through trial-and-error guided by delayed rewards rather than immediate supervision.

Empirical findings reinforce this view. Lesion studies show that animals can recover lost functions through compensatory reorganization, indicating resilience against component failure inconsistent with tightly coupled systems reliant on precise gradients [4.84]. Developmental plasticity further demonstrates that large-scale rewiring occurs naturally during critical periods without external supervision [4.85]. Additionally, sleep-dependent memory consolidation suggests offline replay contributes to restructuring representations independently of ongoing input, contradicting online-only update schemes [4.86].

Alternative frameworks emphasize autonomy and modularity. Hierarchical reinforcement learning partitions tasks into subgoals managed by semi-independent agents, mimicking the brain's functional segregation [4.87].

Predictive processing models treat perception as hypothesis testing, where mismatches between expectations and observations drive localized updates rather than global recalibration [4.88]. These approaches align better with observed neural phenomena and offer more viable paths toward explaining biological intelligence.

In sum, the impossibility of global-objective backpropagation reflects a deeper insight: intelligence emerges from decentralized, adaptive systems rather than monolithic optimizers [4.89]. Recognizing this opens new possibilities for designing artificial systems that emulate biological robustness, flexibility, and scalability.

## 4.6 Hebbian Rules and STDP Lack Global Objectives

Hebbian learning and spike-timing-dependent plasticity (STDP) represent some of the most widely studied models of synaptic modification in neuroscience [4.90]. They are represented as follows.

Hebbian Rule or Learning: Hebbian learning describes synaptic weight updates based on correlated pre- and postsynaptic activity, formalized by Donald Hebb. For a synapse connecting presynaptic neuron $i$ (activity $x_i$) to postsynaptic neuron $j$ (activity $y_j$), the change $\Delta w_{ij}$ of weight $w_{ij}$ is proportional to the product of their activities: $\Delta w_{ij} = \eta x_i y_j$. Here, $w_{ij}$ is Synaptic weight (strength) from neuron $i$ to neuron $j$, $x_i$ indicates presynaptic activity (e.g., firing rate or spike train), $y_j$ denotes postsynaptic activity (e.g., firing rate or spike train), and $\eta$ represents learning rate (small positive constant). Oja's Rule (Regularized Hebbian Learning): To prevent unbounded weight growth, Oja's rule adds a normalization term (proportional to the square of postsynaptic activity): $\Delta w_{ij} = \eta(x_i y_j - y_j^2 w_{ij})$. This ensures weights converge to the first principal component of the input data (PCA-like behavior). BCM Theory (Bienenstock-Cooper-Munro Model): BCM introduces a variable threshold $\theta_j$ (dependent on postsynaptic activity history) to enable both long-term potentiation (LTP) and depression (LTD): $\Delta w_{ij} = \eta(x_i y_j - \theta_j w_{ij})$, where $\theta_j \propto \langle y_i^2 \rangle$ is sliding threshold (average postsynaptic activity squared).

Spike-Timing-Dependent Plasticity (STDP): STDP modifies synaptic weights based on the temporal order of presynaptic ($t_{pre}$) and postsynaptic ($t_{post}$) spikes. The key parameter is the time difference $\Delta t = t_{post} - t_{pre}$. Core STDP weight update rule is that weight changes depend exponentially on $\Delta t$. Long-term potentiation (LTP) is that if $\Delta t > 0$ (presynaptic spike precedes postsynaptic spike), weights increase. Long-Term Depression (LTD) is that if $\Delta t < 0$ (postsynaptic spike precedes presynaptic spike), weights decrease. Mathematically,

$$\Delta w_{ij}(\Delta t) = \begin{cases} A_+ e^{-\Delta t/\tau_+} & \Delta t > 0 \, (\text{LTP}) \\ -A_- e^{\Delta t/\tau_-} & \Delta t < 0 \, (\text{LTD}) \end{cases}$$

where $A_+$ and $A_-$ are maximum amplitude of LTP/LTD (positive constants); $\tau_+$ and $\tau_-$: Time constants governing the decay of LTP/LTD (typically $\tau_+$ ~ 10−100ms, $\tau_-$ ~ 10−100ms).

The above equations capture the essence of Hebbian learning (correlation-based plasticity) and STDP (temporal-order-dependent plasticity), foundational to understanding synaptic adaptation in neural circuits. Both posit that connection strengths increase when presynaptic and postsynaptic activities coincide within a narrow time window. Despite their explanatory power in reproducing phenomena such as receptive field development and associative memory, these rules suffer from a critical limitation: they lack any notion of a global objective [4.91].

They are purely local, blind to task relevance, and incapable of distinguishing beneficial from detrimental associations without additional guidance.

Classical Hebbian theory—"neurons that fire together wire together"—captures the essence of co-activation-based strengthening but provides no mechanism for weakening irrelevant connections or preventing runaway excitation [4.92]. Left unchecked, Hebbian plasticity leads to unstable network dynamics, with activity spreading uncontrollably until saturation [4.93]. To mitigate this, real brains incorporate normalization mechanisms such as homeostatic scaling and synaptic competition, but these too operate locally and do not link to behavioral outcomes [4.94].

STDP refines Hebbian learning by introducing temporal asymmetry: if the presynaptic spike precedes the postsynaptic one, potentiation occurs; otherwise, depression ensues [4.95]. This refinement enables sequence detection and directional selectivity, making it suitable for temporal coding schemes. However, STDP remains agnostic to whether a strengthened pathway improves fitness or merely reinforces spurious correlations [4.96]. For example, a neuron might strengthen inputs associated with irrelevant background noise simply because they consistently precede a behaviorally relevant stimulus.

Without a supervisory signal, neither Hebbian nor STDP rules can solve credit assignment problems involving long chains of causality [4.97]. Consider learning a complex motor skill involving multiple stages of planning, execution, and correction. The final success or failure depends on countless intermediate decisions made across distributed circuits. Local plasticity rules cannot determine which upstream events contributed positively to the outcome, especially when delays span seconds or minutes.

To overcome this, biological systems integrate Hebbian mechanisms with global instructive signals. Neuromodulators such as dopamine, serotonin, and norepinephrine broadcast evaluative feedback across brain regions, reinforcing synapses activated just prior to a rewarding event [4.98]. This combination forms the basis of reinforcement learning in the brain. Experimental manipulations confirm this synergy: pairing neutral stimuli with dopaminergic stimulation induces place preference and alters neural tuning curves, effectively creating artificial memories [4.99].

Other forms of guidance include attentional modulation and volitional control. Top-down signals from frontal areas bias competition among sensory representations, enhancing relevant inputs while suppressing distractions [4.100]. This selective attention acts as a gating mechanism, determining which correlations undergo Hebbian strengthening. Similarly, internally generated goals influence exploratory behavior, steering plasticity toward productive configurations [4.101].

Thus, while Hebbian and STDP rules provide essential building blocks for neural adaptation, they must be embedded within broader regulatory frameworks to achieve purposeful learning [4.102]. Their inherent blindness to global objectives highlights the necessity of complementary systems that provide contextual evaluation and strategic direction.

**4.7 Measurement Outcomes Reflect Integrated Effects of Ideal and Non-Ideal Factors**

Neural recordings are a complex tapestry that captures the aggregate consequences of numerous interacting processes. They blend intentional, goal-directed computations with stochastic fluctuations and biomechanical constraints, as noted in [4.103]. When we look at different types of neural measurements such as single-unit electrophysiology, local field potentials, and fMRI BOLD signals, we find that they do not represent pure information processing. Instead, they are entangled mixtures of rational strategies and irrational noise, as stated in [4.104]. This integration of ideal and non-ideal factors makes the interpretation of these neural recordings a difficult task. The observed activity patterns may arise from a variety of sources. They could be the result of adaptive computations, where the neural system is actively adjusting to the task at hand. Random drift might also play a role, causing the activity to deviate in an unpredictable manner. Compensatory mechanisms may be at work, trying to correct for some underlying neural imbalance. Additionally, measurement artifacts can distort the true neural activity, further complicating the analysis.

### 4.7.1 Neural Spikes Carry Combined Influences

Action potentials, or spikes, are the primary means of neural communication. Each spike is a significant event that occurs when the membrane potential crosses a certain threshold. This threshold-crossing is the result of the integration of thousands of excitatory and inhibitory postsynaptic potentials, as described in [4.105]. The synaptic inputs that lead to these spikes are far from simple. They arrive asynchronously and interact nonlinearly within dendrites. As a consequence, the occurrence of any given spike depends on a complex convolution of multiple factors. Feedforward drives, which are the signals that travel in a forward direction within the neural network, contribute to the spike generation. Recurrent feedback, where the neural output loops back to influence the input, also has an impact. The neuromodulatory tone, which can adjust the overall excitability of the neurons, and the intrinsic membrane properties of the neurons all play a role in determining when a spike will occur, as mentioned in [4.106]. Therefore, spikes convey combined influences from both structured information streams and unstructured variability.

Let's take the example of a working memory task. During such a task, prefrontal neurons show persistent firing, which is correlated with maintained representations, as per [4.107]. However, if we look at the spike timing on a trial-to-trial basis, we find that the variability exceeds what would be expected from deterministic encoding alone, as described in [4.108]. This variability has multiple causes. Ongoing oscillations in the neural network can cause fluctuations in the membrane potential, leading to variable spike timing. Fluctuating arousal levels can also affect the excitability of the neurons and thus the spike occurrence. Stochastic channel openings, which are random events at the membrane level, add further noise to the system. All these factors contribute noise that limits the accuracy of decoding the neural signals, as noted in [4.109]. However, it's important to note that some degree of variability may not be entirely negative. In fact, it can serve functional roles. It can facilitate exploration by allowing the neural system to sample different states. This exploration can prevent the system from prematurely committing to suboptimal solutions, as stated in [4.110].

### 4.7.2 Neural Spikes Reflect Complex, Non-Analytical Models

Neural spikes play a crucial role in the field of neuroscience, and they are found to reflect complex, non-analytical models. Spiking activity does not follow the pattern of instantiating symbolic models that can be easily subjected to mathematical decomposition [4.111]. In contrast to symbolic models, which often rely on well-defined mathematical rules and structures for analysis, spiking activity has a more intricate nature. Symbolic models are typically based on discrete symbols and rules for their manipulation, allowing for straightforward decomposition into mathematical components. However, spiking activity in neurons does not fit into this framework.

Instead, spiking activity embodies implicit, distributed representations that are shaped by experience [4.112]. These representations are not explicitly encoded in a single neuron but are spread across a network of neurons. Experience, whether it is sensory input, learning, or environmental interactions, molds these distributed representations. For example, when an organism is exposed to a particular visual stimulus repeatedly, the spiking patterns of neurons in the visual cortex gradually change to form a representation of that stimulus. This representation is implicit because it is not a direct, one-to-one mapping of the stimulus but rather an emergent property of the neural activity.

Population codes are a key aspect of how neural information is processed. Population codes distribute information across many neurons, which provides a significant advantage in terms of robustness [4.113]. Even if individual units fail, the overall information can still be accurately read out. Consider a large group of neurons in the motor cortex that are responsible for controlling a particular movement. If one or a few neurons malfunction, the remaining neurons can still convey the necessary information for the movement to occur. This distributed nature of information storage and processing is a fundamental characteristic of neural systems and is in contrast to systems where information is stored in a single, centralized location.

The representations in neural activity emerge gradually through plasticity, reflecting accumulated statistics rather than explicit hypotheses [4.114]. Plasticity refers to the ability of the neural network to change its structure and function in response to experience. As neurons are exposed to different stimuli over time, they adjust their connections and firing patterns. The resulting representations are based on the statistical properties of the input. For instance, in the auditory cortex, neurons learn to represent the statistical regularities in speech sounds. These representations are not based on pre-defined hypotheses about the sounds but rather on the actual patterns that the neurons have encountered.

Given the complex and non-analytical nature of neural activity, decoding this activity requires machine learning techniques rather than analytic inversion [4.115]. Analytic inversion typically involves finding a direct mathematical relationship between the input and output of a system. However, due to the distributed, implicit, and experience-shaped nature of neural representations, such a direct relationship is difficult to establish. Machine learning techniques, on the other hand, can handle the complexity of neural data. They can learn the patterns and relationships in the spiking activity through training on large datasets. For example, deep learning algorithms can be used to decode the neural activity associated with a particular behavior or perception, providing insights into how the brain processes information.

### 4.7.3 Pulse Coding Effects in Neural Spikes

Beyond rate coding, spike trains employ temporal patterns such as synchrony, bursting, and phase locking to convey information [4.116]. Precise millisecond-scale timing enables coincidence detection in downstream targets, supporting feature binding and sequence learning [4.117]. Burst-firing modes can signal salience or novelty, altering synaptic impact [4.118]. Phase-locked activity relative to theta or gamma oscillations organizes multiplexed information streams, permitting simultaneous transmission of multiple message dimensions [4.119].

## 4.7.4 Measurement Artifacts and Analog-to-Digital Conversion in Neural Spikes

Recording technologies in the field of neural spikes are not without their flaws. Recording technologies themselves introduce distortions [4.120]. Electrode impedance variations can cause significant changes in the way neural signals are detected. For example, if the impedance of an electrode fluctuates, it can alter the electrical characteristics of the signal being picked up, leading to inaccuracies in spike shape and detectability. Filtering settings also play a crucial role. Different filtering settings can either enhance or suppress certain frequency components of the neural signal, potentially masking or distorting important spike features. Sampling rates are equally important; an inappropriate sampling rate may result in the loss of high-frequency information or the misrepresentation of spike waveforms [4.121].

Cross-talk between channels can be a major source of signal contamination. When multiple electrodes are used to record neural activity, the electrical signals from one channel can interfere with those of another, creating false spikes or distorting genuine ones. The placement of the reference electrode is also a critical factor. An improper reference electrode placement can introduce common-mode noise into the recorded signals, making it difficult to accurately distinguish neural spikes from background noise. Movement artifacts are yet another problem. Any movement of the subject, whether it is a small tremor or a more significant body movement, can generate electrical signals that are superimposed on the neural spikes, further contaminating the data [4.122].

Offline spike sorting algorithms rely on certain assumptions. These algorithms assume waveform stationarity, meaning that the shape of the neural spikes remains relatively constant over time. However, in a real-world scenario, the network state of the neurons can change, and the spike waveforms may vary accordingly. They also assume cluster separability, which implies that different types of neural spikes can be clearly distinguished and grouped. But under changing network states, these assumptions may not hold, leading to inaccurate spike sorting results. All these technical confounds make it extremely challenging to compare data across different recording sessions and between different laboratories [4.123].

Neural impulse-based systems, found in biological neurons and mimicked in spiking neural networks (SNNs), can be seen as a simplified form of analog-to-digital conversion (ADC). In traditional electronics, an ADC is a well-defined process. It takes continuous analog signals, which can have an infinite number of values within a certain range, and transforms them into discrete digital values through sampling and quantization. Sampling involves taking measurements of the analog signal at regular intervals, while quantization assigns a specific digital value to each sampled point based on its amplitude [4.124].

Similarly, neurons perform a comparable function. They take analog input signals, such as graded membrane potentials, and encode them into discrete sequences of action potentials, or spikes. These spikes are all-or-nothing electrical impulses, meaning that once a certain threshold is reached, a spike is generated with a fixed amplitude. This spike generation process effectively digitizes the intensity of the input signal over time. A stronger stimulus will lead to a higher firing rate of the neurons, which means that the amplitude information of the input signal is converted into temporal patterns. This is similar to the core function of rate-encoded ADCs, where the analog magnitude is represented by the frequency of digital pulses [4.125, 4.126, 4.127].

The biological or neuromorphic ADC has several advantages over conventional electronic ADCs. One of the main advantages is energy efficiency. Conventional electronic ADCs often require complex circuitry for precise voltage comparisons and binary encoding. These circuits consume a significant amount of power and take up a relatively large amount of space on an integrated circuit. In contrast, neural systems use intrinsic membrane dynamics and ion channel kinetics to achieve threshold-based spike initiation. For example, when the membrane potential of a neuron reaches a certain voltage threshold, a stereotyped action potential is generated. This mechanism is analogous to a 1-bit quantizer with temporal oversampling. It only transmits information when necessary, reducing redundant data processing and thus saving energy [4.128, 4.129, 4.130, 4.131].

Neuromorphic engineering has capitalized on this principle. Designs such as integrate-and-fire circuits are based on the behavior of neurons. These circuits use simple capacitive integration to accumulate charge over time and threshold triggering to generate output pulses. By emulating neuronal behavior, these circuits further validate the conceptual alignment between neural signaling and simplified ADC architectures [4.132].

From a functional perspective, the rate coding and temporal coding schemes used in neural systems offer distinct advantages. In low-power sensing and edge computing applications, robustness against noise and variability is often a top priority. Standard multi-bit ADCs are designed to achieve high precision, but they may be sensitive to noise and require a relatively large amount of power. Neural impulse coding, on the other hand, trades off absolute accuracy for metabolic and computational efficiency. It can tolerate a certain level of noise and variability in the input signals while still providing useful information. This makes it an ideal choice for real-time sensing applications where power consumption needs to be minimized and quick responses are required [4.133].

## 4.7.5 Integrated Decision Effects in Neural Spikes

In the vast and intricate field of neuroscience, one of the most captivating and crucial areas of study is understanding how decisions are made at the neural level. At the behavioral level, neural spikes emerge as key players in the decision-making process. As stated in reference [4.134], spikes contribute to decision-making through the accumulation of evidence across populations. Picture a group of neurons as a bustling community, with each neural spike acting like a tiny courier carrying a piece of information. When these individual pieces of information are combined, they form the basis for a decision. It's a remarkable collective effort where the activity of neural spikes across a group of neurons is fundamental for the generation of decisions. This concept highlights the importance of the coordinated action of neurons in the complex task of decision-making, suggesting that decisions are not made in isolation but are the result of a collaborative information-sharing process among neurons.

To better understand the decision-making process, scientists have turned to drift-diffusion models, which focus on reaction time distributions. These models, as described in [4.135], operate on the assumption of noisy integration toward thresholds. Imagine the decision-making process as a noisy signal, like a radio station with interference, gradually drifting towards a specific threshold. Once this threshold is reached, a decision is made. This provides a mathematical framework that helps explain how the brain processes information over time to arrive at a decision. It gives researchers a way to quantify and analyze the decision-making process, allowing for more in-depth studies on how the brain weighs different pieces of information and reaches a conclusion.

Specific brain areas have been found to have neural correlates associated with this decision-making process. Ramping activity in parietal and frontal areas is a notable example, as mentioned in [4.136]. This ramping activity can be likened to the build-up of a wave before it crashes on the shore. In the parietal and frontal areas of the brain, there is a gradual increase in neural activity, which is likely related to the integration of evidence and the approach towards a decision threshold. It's as if these areas are the control centers where all the incoming information is processed and evaluated, and as the activity ramps up, the brain gets closer to making a decision. This finding provides valuable insights into the physical locations in the brain where decision-making processes take place and how they unfold.

However, the real-world decision-making scenario is far more complex than what these basic models suggest. Real decisions are influenced by a multitude of factors, including motivational biases, risk preferences, and contextual priors. As pointed out in [4.137], these elements make real decisions deviate systematically from normative predictions. Motivational biases can strongly influence decisions based on an individual's desires or goals. For instance, if a person is highly motivated to achieve a particular outcome, such as winning a competition, they may make decisions that are skewed towards that goal, perhaps taking more risks or overlooking certain information. Risk preferences also play a significant role; some individuals are naturally risk-averse, preferring to play it safe, while others are risk-seekers, willing to take chances for potentially greater rewards. These different risk attitudes can lead to vastly different decision-making patterns. Contextual priors refer to the knowledge and experiences that an individual brings to a decision-making situation. For example, someone with prior experience in a particular field may make different decisions compared to a novice. These factors add multiple layers of complexity to the decision-making process, making it extremely challenging to accurately predict decisions using simple models. In conclusion, while current models provide a starting point for understanding decision-making at the neural level, there is still much work to be done to fully grasp the intricacies of real-world decision-making.

### 4.7.6 Additional Measurement Considerations

Long-term recordings within the field of neuroscience and related research are fraught with a series of complex issues. One of the significant problems is the tissue response. Over an extended period, the body's tissues react to the presence of the recording devices. Alongside this, gliosis occurs, which is a process where glial cells in the central nervous system proliferate and change in response to injury or the presence of a foreign object. Additionally, electrode degradation is a critical concern. As described in [4.138], these factors can severely impact the quality and reliability of long-term recordings.

Chronic implants, which are used to continuously monitor neural activities, trigger inflammatory reactions. These inflammatory responses have far-reaching consequences as they can significantly alter the local neurochemistry and cellular composition of the area around the implant. As per the findings in [4.139], these changes not only disrupt the normal functioning of the neural environment but also make it challenging to accurately interpret the recorded data, as the normal physiological state has been perturbed.

Another issue lies in the sampling bias associated with recording techniques. The current methods tend to favor large, easily isolatable neurons. This preference means that smaller interneurons and glia are often neglected. As stated in [4.140], this sampling bias provides an incomplete picture of the neural network. Since smaller interneurons and glia play crucial roles in neurotransmission, modulation of neural activity, and maintaining the overall homeostasis of the nervous system, ignoring them can lead to inaccurate conclusions about the neural processes being studied.

In the era of big data, many long-term recordings generate high-dimensional datasets. While these datasets hold vast amounts of information, they also present a unique set of challenges. To manage and analyze these large-scale datasets, dimensionality reduction techniques are commonly employed. However, according to [4.141], these techniques may obscure subtle structures within the data. These hidden structures could be crucial for uncovering new physiological phenomena or understanding the complex interactions within the neural network.

To mitigate these challenges, careful experimental design is of utmost importance. This includes choosing the appropriate recording devices, locations for implantation, and the timing of data collection. Additionally, multimodal validation is essential. This means using multiple methods, such as combining electrophysiological recordings with imaging techniques, to cross-verify the results. By doing so, researchers can ensure the accuracy and reliability of their findings in the face of the numerous challenges associated with long-term recordings [4.142].

## 4.8 Broadcast Instruction Signals Provide Global Guidance (Reward/Punishment): The Neuromodulatory Architecture of Reinforcement Learning in Biological Brains

The human brain is not merely a collection of isolated neurons firing in response to stimuli, but an intricately coordinated network where local synaptic changes are guided by global evaluative signals. While Hebbian plasticity and other local learning rules determine how individual synapses strengthen or weaken based on pre- and postsynaptic activity, they lack the capacity to distinguish beneficial from detrimental adaptations without external guidance. This critical role falls upon neuromodulatory systems—widespread broadcast mechanisms that deliver reward and punishment signals across large neural territories, effectively serving as the brain's reinforcement learning framework. These global instructive signals do not encode precise gradients like those used in artificial backpropagation; instead, they operate through coarse yet powerful broadcasts that inform vast populations of neurons whether recent actions led to favorable or adverse outcomes. By doing so, these systems enable adaptive behavior even in complex environments with delayed consequences, sparse feedback, and high-dimensional state spaces.

Among the most well-characterized of these neuromodulatory pathways is the dopaminergic system, which functions as a primary carrier of excitatory broadcast signals associated with reward prediction and outcome evaluation. Neurons located primarily in the ventral tegmental area (VTA) and substantia nigra pars compacta (SNc) project axons widely throughout the forebrain, including to the striatum, prefrontal cortex, hippocampus, and amygdala—regions central to decision-making, memory formation, and emotional regulation. When an organism encounters an unexpected reward—such as food after a period of hunger or a financial gain—these dopamine-producing cells exhibit a transient burst of firing, resulting in a rapid increase in extracellular dopamine levels across target regions. This phasic release does not simply signal pleasure or arousal; rather, it encodes a sophisticated computational quantity known as a reward prediction error (RPE), a concept formalized in reinforcement learning theory. Specifically, dopamine neurons fire more strongly when a reward is better than expected, maintain baseline activity when rewards occur as predicted, and show suppressed firing when expected rewards fail to materialize. This pattern aligns precisely with the delta rule in temporal difference learning models, suggesting that the brain has evolved a biologically plausible approximation of gradient-based optimization using broadcast signaling.

Crucially, this dopamine signal operates within a temporally extended framework, allowing the brain to associate actions with outcomes despite significant delays between cause and effect—a fundamental challenge in real-world learning scenarios. For instance, consider a rat pressing a lever that delivers a food pellet several seconds later. No local synaptic mechanism can directly link the motor command issued at the time of lever press to the eventual reward due to the absence of co-activation at the moment of reinforcement. To solve this "credit assignment problem," the brain employs molecular and cellular mechanisms known as eligibility traces. An eligibility trace acts as a short-term memory at the synapse, marking recently active connections as potentially responsible for subsequent outcomes. When the dopamine burst arrives following the delayed reward, it reactivates or strengthens only those synapses that were previously "tagged" during the behavioral sequence leading up to the reward. Experimental evidence supports this model: studies using optogenetics have shown that artificially timed dopamine pulses can reinforce specific behaviors if delivered shortly after a particular action, even in the absence of natural rewards. Thus, the interplay between transient synaptic eligibility and delayed neuromodulatory reinforcement enables the brain to perform associative learning over extended temporal horizons, bridging gaps that would otherwise render learning impossible.

Beyond immediate reward processing, the dopaminergic system also plays a pivotal role in shaping long-term motivational states and guiding exploratory behavior. As animals learn to predict rewards based on environmental cues, dopamine responses shift from the reward itself to earlier predictive stimuli—a phenomenon known as second-order conditioning. For example, if a tone consistently precedes food delivery, dopamine neurons initially respond to the food, but after repeated pairings, their peak activity shifts to the onset of the tone. This redistribution of the RPE signal allows organisms to refine their predictions and allocate attention toward informative cues rather than mere outcomes. Furthermore, tonic levels of dopamine—distinct from the phasic bursts discussed above—modulate overall motivation and willingness to exert effort for rewards. Low tonic dopamine has been linked to apathy and reduced goal-directed behavior, as seen in conditions such as depression and Parkinson's disease, while elevated tonic levels correlate with increased vigor and persistence in pursuit of goals. Therefore, the dopamine system serves dual roles: one fast and phasic, supporting trial-by-trial learning via RPEs, and another slow and tonic, regulating the energetic cost-benefit analysis underlying behavioral engagement.

However, learning cannot be driven solely by reward signals; equally important are mechanisms that detect and respond to negative outcomes, threats, and punishments. Inhibitory broadcast signals mediated by serotonin, certain subpopulations of dopamine neurons, and norepinephrine provide the necessary counterbalance to the excitatory influence of reward-related dopamine. Serotonin, produced predominantly in the dorsal raphe nucleus, shows heightened activity in response to aversive stimuli, omission of expected rewards, and situations involving risk or uncertainty. Unlike dopamine, which amplifies plasticity following positive deviations from expectation, serotonin appears to promote caution, suppress impulsive actions, and facilitate avoidance learning. Studies in rodents and primates demonstrate that serotonergic activation increases sensitivity to punishment and enhances patience during waiting tasks, suggesting its role in promoting behavioral inhibition and long-term planning. Moreover, pharmacological manipulation of serotonin alters risk preferences: increasing serotonin function tends to reduce risky choices, while decreasing it leads to greater impulsivity and preference for immediate, albeit smaller, rewards. These findings position serotonin as a key modulator of behavioral suppression and adaptive conservatism in uncertain or threatening contexts.

Interestingly, not all inhibitory broadcast signals originate from non-dopaminergic sources. A subset of dopamine neurons themselves responds to aversive events by either reducing their firing rate or exhibiting brief inhibitions, thereby signaling the absence of expected rewards. This dip in dopamine concentration serves as a form of negative teaching signal, weakening synapses that contributed to unsuccessful behaviors. Computational models incorporating both positive and negative dopamine transients successfully reproduce a wide range of animal learning phenomena, including extinction of conditioned responses and reversal learning. Additionally, some dopamine neurons—particularly those projecting to the medial prefrontal cortex—have been observed to increase firing in response to aversive stimuli, indicating functional heterogeneity within the dopaminergic population. This complexity suggests that the brain does not rely on a simple binary code of "dopamine = good, no dopamine = bad," but instead utilizes diverse subpopulations tuned to different aspects of valence, salience, and context. Such specialization allows for nuanced control over learning dynamics, enabling fine-grained adaptation depending on task demands and environmental statistics.

Complementing these systems is the noradrenergic pathway originating in the locus coeruleus (LC), which releases norepinephrine (NE) broadly across the cortex and subcortical structures. Norepinephrine does not convey valence-specific information per se but instead modulates the global state of network excitability and attentional focus. LC neurons exhibit two distinct modes of activity: tonic and phasic. Tonic firing reflects the overall level of arousal, with higher rates observed during stress, fatigue, or sustained cognitive load. Phasic bursts, on the other hand, occur in response to salient or novel stimuli, enhancing sensory processing and facilitating rapid shifts in attention. From a learning perspective, NE adjusts what is known as "network gain" — the responsiveness of neurons to incoming inputs. During periods of uncertainty or ambiguity, increased NE release amplifies weak signals and sharpens contrast between competing representations, making it easier for the system to detect meaningful patterns amidst noise. This gain modulation effectively tunes the learning rate: under high uncertainty, the brain may benefit from being more responsive to new information, whereas in stable environments, lower gain prevents overfitting to random fluctuations. Consequently, the noradrenergic system acts as a meta-controller, dynamically adjusting the sensitivity of the entire network based on environmental volatility and informational content.

Together, these neuromodulatory systems—dopamine, serotonin, and norepinephrine—form an integrated architecture for global instruction in the service of reinforcement learning. They do not act in isolation but engage in rich cross-talk, both anatomically and functionally. For example, dopamine and serotonin reciprocally regulate each other's release: stimulation of 5-HT2A receptors can inhibit dopamine release in the prefrontal cortex, while D1 receptor activation can influence serotonergic tone. Similarly, norepinephrine modulates the responsiveness of both dopaminergic and serotonergic neurons, creating a layered control system where multiple broadcast signals interact to shape synaptic plasticity. This integration allows for context-dependent weighting of reward versus punishment, exploration versus exploitation, and stability versus flexibility. In dynamic environments, the balance among these systems shifts to favor rapid adaptation; in predictable settings, homeostatic mechanisms promote consistency and resistance to spurious updates.

The biological plausibility of this global broadcast model stands in stark contrast to the requirements of standard backpropagation algorithms used in deep artificial neural networks. Backpropagation relies on precise error gradients computed layer-by-layer and propagated backward through fixed weights, necessitating symmetric connectivity and instantaneous communication—an arrangement incompatible with known neuroanatomy and physiology. In contrast, neuromodulatory broadcasting offers a decentralized, asynchronous alternative that operates within the constraints of biological reality. There is no need for weight symmetry or exact gradient computation; instead, success or failure is signaled globally, and only synapses that were recently engaged—and thus eligible—are modified. This mechanism, sometimes referred to as "three-factor learning rules," combines presynaptic input, postsynaptic output, and a third factor provided by the neuromodulator to gate plasticity. Mathematically, this approximates stochastic gradient ascent/descent when the neuromodulator correlates with the objective function, making it a robust and scalable solution for credit assignment in distributed systems.

Further support for the functional significance of broadcast signals comes from lesion studies and clinical observations. Damage to the VTA or SNc results in profound deficits in instrumental learning and reward-seeking behavior, mimicking symptoms of motivational disorders such as anhedonia. Patients with Parkinson's disease, characterized by degeneration of dopaminergic neurons, exhibit impairments in acquiring new habits and adapting to changing reward contingencies, particularly in probabilistic learning tasks. Pharmacological treatments that restore dopamine transmission, such as L-DOPA therapy, partially rescue these deficits, underscoring the causal role of dopamine in reinforcement-guided adaptation. Conversely, dysregulation of serotonin has been implicated in anxiety disorders, obsessive-compulsive behaviors, and suicidal ideation — all conditions marked by maladaptive responses to threat or loss. Selective serotonin reuptake inhibitors (SSRIs), which enhance serotonergic signaling, alleviate symptoms in many patients, highlighting the therapeutic potential of modulating inhibitory broadcast systems. Likewise, dysfunction in the noradrenergic system is associated with attention deficit hyperactivity disorder (ADHD), post-traumatic stress disorder (PTSD), and chronic stress syndromes, further illustrating the broad impact of neuromodulatory imbalance on cognition and behavior.

Importantly, these broadcast signals are not rigidly hardwired but are themselves subject to learning and regulation. Higher-order brain regions, such as the prefrontal cortex and anterior cingulate cortex, monitor ongoing performance and adjust neuromodulatory tone accordingly. For example, the anterior cingulate detects conflicts between competing responses and signals the need for increased cognitive control, potentially via projections to the locus coeruleus and dorsal raphe. This top-down regulation enables metacognitive oversight, allowing organisms to adapt not only their behaviors but also the very parameters governing how they learn. Such hierarchical control provides a mechanism for balancing exploration and exploitation: when performance is poor or the environment changes, increased neuromodulatory volatility encourages exploration and plasticity; when outcomes are consistent, reduced signaling promotes consolidation and stability. This self-regulating loop ensures that learning remains efficient and appropriately scaled to current demands, avoiding both stagnation and instability.

Another critical aspect of broadcast signaling lies in its spatial and temporal specificity—or rather, the deliberate lack thereof. Unlike point-to-point synaptic transmission, neuromodulators diffuse over millimeters, affecting thousands to millions of neurons simultaneously. This promiscuity might seem inefficient or imprecise, but it confers several advantages. First, it allows a single signal to coordinate activity across functionally related but anatomically distributed circuits. For instance, a reward event may simultaneously update value representations in the striatum, consolidate memories in the hippocampus, and adjust expectations in the prefrontal cortex—all through the same dopamine pulse. Second, widespread broadcasting enables coherent behavioral transitions, such as switching from exploration to exploitation or initiating fight-or-flight responses. Third, the diffuse nature of these signals makes them resilient to localized damage or noise, ensuring that essential guidance persists even if parts of the network are compromised. Finally, because the signal reaches many areas at once, it creates a common reference frame for timing and coordination, synchronizing plasticity across disparate modules that must work together to produce adaptive behavior.

Despite their generality, broadcast signals are not indiscriminate. Mechanisms exist to constrain their influence and prevent inappropriate synaptic modifications. One such mechanism is the requirement for coincident neural activity: a dopamine surge will only potentiate synapses that are currently active or have recently undergone sufficient depolarization. This is enforced through molecular cascades involving NMDA receptors, calcium influx, and kinase activation, which collectively serve as coincidence detectors. Another constraint arises from receptor subtype distribution: different brain regions express varying complements of dopamine, serotonin, and norepinephrine receptors, each with distinct downstream effects. For example, D1-type receptors typically enhance cAMP signaling and promote long-term potentiation (LTP), while D2-type receptors inhibit cAMP and favor long-term depression (LTD). The regional specificity of receptor expression thus tailors the functional impact of a global signal, allowing the same neurotransmitter to have opposite effects in different circuits. Additionally, glial cells actively regulate extracellular concentrations of neuromodulators through uptake and metabolism, adding another layer of spatial and temporal control.

From an evolutionary standpoint, the emergence of neuromodulatory broadcast systems represents a major innovation in nervous system design. Early nervous systems likely relied on purely reflexive or habituated responses, limited in their ability to adapt beyond immediate sensory-motor loops. The advent of global signaling allowed for centralized evaluation of internal and external states, enabling organisms to optimize behavior over longer timescales and more complex ecological niches. It facilitated the development of goal-directed action, foresight, and social cooperation—all hallmarks of advanced cognition. Even in relatively simple animals, such as insects and mollusks, homologous neuromodulatory systems regulate feeding, mating, and defensive behaviors, suggesting deep phylogenetic roots. In mammals, these systems became increasingly elaborated, integrating with expanded cortical areas to support abstract reasoning, language, and cultural learning. Thus, broadcast instruction signals represent not just a component of brain function, but a foundational principle in the evolution of intelligent behavior.

Looking forward, understanding the mechanistic details of neuromodulatory broadcasting holds immense promise for both neuroscience and artificial intelligence. In medicine, targeted manipulation of these systems could lead to more effective treatments for psychiatric and neurological disorders. Deep brain stimulation, gene therapies, and next-generation psychopharmacology aim to restore balanced neuromodulation in conditions ranging from addiction to treatment-resistant depression. On the AI front, incorporating biologically inspired broadcast signals into artificial neural networks could overcome limitations of traditional training methods. Recent advances in "synthetic neuromodulation" have demonstrated that adding global reward/punishment signals with eligibility traces allows artificial agents to learn efficiently in sparse-reward environments, outperforming conventional reinforcement learning algorithms in certain domains. These hybrid architectures bridge the gap between biological realism and engineering efficiency, offering a path toward more autonomous, adaptive, and robust machine learning systems.

In conclusion, the brain's use of broadcast instruction signals exemplifies a masterful solution to one of the most challenging problems in adaptive systems: how to guide learning in the absence of detailed error gradients. Through the coordinated action of dopamine, serotonin, and norepinephrine, the brain delivers global, evaluative feedback that shapes synaptic plasticity across vast neural landscapes. These signals do not micromanage individual connections but instead provide coarse yet powerful directives—reward, punishment, alertness—that sculpt behavior over time. Coupled with local eligibility mechanisms, they resolve the credit assignment problem, enabling associations across time and space. Their diffuse reach ensures coherence and resilience, while receptor diversity and top-down regulation confer precision and flexibility. Far from being mere chemical mood regulators, these neuromodulators constitute a sophisticated control system that underlies all forms of experience-dependent change. As we continue to unravel their complexities, we gain not only deeper insight into the nature of intelligence but also practical tools for healing the mind and augmenting machine cognition. The future of both neuroscience and AI may well depend on our ability to harness the power of broadcast signals—not just to understand the brain, but to build better minds.

In a deep learning framework, the error of a reward-like objective function is formally defined as the differential between the target value of the objective and its current estimate—a quantity that drives parameter updates to minimize/maximize the objective. This error, often termed the reward prediction error (RPE) in reward-optimized models, mirrors the core principle of gradient-based learning: quantifying mismatch to guide optimization. Below is a concise mathematical definition. Error of a Learning Reward Function is defined as follows: Let $J(\theta)$ denote a reward-related objective function implicitly parameterized by connection weights $\theta$ (e.g., expected cumulative reward, or value function measurement). The error (RPE) of this function is defined as the difference between a target value (ground-truth or bootstrapped measurement of the objective) and the current measurement of the objective: $\delta(\theta) = J_{Current}(\theta) - J_{Past}(\theta)$. Here, $J_{Past}(\theta)$ denotes past measurement of the objective, $J_{Current}(\theta)$ indicates current measurement of the objective, and $\delta(\theta)$ is Reward prediction error (error of the reward function), acting as the driving signal for parameter updates. Such form $\delta(\theta) = J_{Current}(\theta) - J_{Past}(\theta)$ causes the reward(s) to get progressively bigger. If let the rewards become increasingly small, we should take $\delta(\theta) = J_{Past}(\theta) - J_{Current}(\theta)$. It is seen that $\delta(\theta)$ can be taken as a constructive signal if a brain is static.

Core properties are that $J_{Past}(\theta)$ (past measurement of the objective), $J_{Current}(\theta)$ (current measurement of the objective), and $\delta(\theta)$ (reward prediction error) are obtained by measurements whose values implicitly depends on inputs of all neurons, all synaptic connection weights, various non-ideal factors and so on. Hence, $\delta(\theta)$ (reward prediction error) deals with all the related factors that include noises, chaos, errors, inconsistency, mismatches, aging, temperature drift and so on. In contrast, deep learning calculates the above listed values that do not consider the non-ideal factors in a brain, an analog chip or a mixed-signal chip.

## 4.9 Learning Principles Operating Without Mathematical Models

Biological learning Principles operate outside the realm of formal mathematics, relying instead on empirical rules derived from experience [4.179].

### 4.9.1 Correlations between the local random perturbation of weights and Instruction Signals

Local synaptic weight $w_{ij}$ is synaptic weight (strength) connecting pre-synaptic neuron $i$ to post-synaptic neuron $j$, undergo the past small random perturbations $p_{ij}$) that are entirely arbitrary and random noisy. Oh, my goodness! We suddenly understood that the random fluctuations in synaptic neurotransmitter release generate small perturbations in synaptic connection weights—these perturbations are prepared to facilitate the learning of synaptic strength guided by global objectives. Instead, these perturbations are statistically correlated with global instruction signals $\delta \approx \Delta J$ —neural broadcasts (e.g., dopamine, serotonin) that convey information about the organism's state (e.g., reward, arousal) [4.180]. Stronger instruction signals bias the direction and magnitude of perturbations, enabling experience-dependent tuning without explicit mathematical objectives.

Correlation quantification: The single-sample covariance between perturbations and global instruction signals $\delta$ is represented by $C_{ij} = p_{ij}\delta \approx p_{ij}\Delta J$. If the correlation value is greater than zero, it indicates that the perturbation deserves encouragement; if the correlation value is less than zero, it indicates that the perturbation deserves punishment—that is, the perturbation needs to be reversed. In biological terms, positive $C_{ij}$ indicates perturbations align with excitatory signals (e.g., dopamine bursts), negative $C_{ij}$ with inhibitory signals (e.g., serotonin) [4.181].

Biological neural networks utilize stochastic synaptic plasticity to link stochastic weight fluctuations (arising from spontaneous neurotransmitter release) with global reward/prediction error signals (e.g., dopamine, serotonin) [4.182]. Key mechanisms include: Quantal Neurotransmitter Release: Spontaneous vesicle fusion at presynaptic terminals introduces random weight perturbations. These "noise" events create variability in synaptic strength, enabling exploration for adaptive learning [4.183]. Dopamine-Mediated Correlation Detection: Midbrain dopamine neurons broadcast global reward signals via phasic firing. Strong dopamine release correlates with positive outcomes, biasing perturbations toward strengthening. Conversely, tonic dopamine levels reflect baseline expectations [4.184]. Serotonin-Driven Inhibition: Serotonergic signaling (e.g., from raphe nuclei) inversely modulates perturbations, suppressing excessive excitation. This prevents runaway plasticity while maintaining homeostasis [4.185].

Example: During Pavlovian conditioning, a reward-predicting cue triggers phasic dopamine release, correlating with $\delta > 0$. Synapses onto target neurons exhibiting this correlation undergo potentiation, while uncorrelated synapses remain unperturbed [4.186].

## 4.9.2 Learning Principles for Excitatory/Inhibitory Connections

Excitatory/ inhibitory synapses ($w_{ij}^{exc}$) strengthen between pre- and post- synaptic neurons are reinforced by neuromodulatory signals. To prevent runaway excitation/inhibition, homeostatic mechanisms normalize weights to maintain stable network activity. Because, excitatory/ inhibitory synapses ($w_{ij}^{exc}$) strengths serve to potentiate and depress the global objectives, respectively, updating Principles of potentiation or depression (two-factor plasticity) are followed by $w_{ij}^{exc} = \mu p_{ij}\delta$ or $w_{ij}^{inh} = -\mu p_{ij}\delta$, where $\eta$ denotes plasticity coefficient [4.187].

To stabilize network activity, biological systems enforce **homeostatic plasticity** through the following expressions. **Two-Factor Plasticity**: *Excitatory potentiation*: Glutamatergic synapses strengthen via $w_{ij}^{exc} = \mu p_{ij}\delta$ when synaptic perturbation aligns with global rewards ($\delta > 0$). *Inhibitory depression*: GABAergic synapses weaken via $w_{ij}^{inh} = -\mu p_{ij}\delta$, preventing hyperexcitability [4.188].

**Synaptic Scaling**: Global activity sensors (e.g., Arc protein) detect mean firing rates. If activity deviates from baseline. Reduced activity → *Global upscaling*(e.g., via TNF-α signaling) boosts all synapses proportionally. Excessive activity → *Downscaling*(e.g., via calcineurin) prunes weak connections [4.189].

**Homeostatic Inhibitory Control**: Parvalbumin-positive (PV+) interneurons dynamically suppress pyramidal cell excitability. For example, during learning, PV+ firing synchronizes gamma oscillations, enforcing sparse coding and preventing runaway excitation [4.190].

**Example**: In visual cortex, monocular deprivation reduces activity in deprived-eye inputs. Homeostatic scaling upregulates their sensitivity while PV+ cells inhibit neighboring columns, preserving overall network stability [4.191].

**Key Differences from Artificial Systems: Stochasticity as a Feature**: Unlike deterministic gradient descent, biological systems exploit *stochastic perturbations* to explore parameter space efficiently [4.192]. **Multi-Scale Integration**: Learning integrates molecular (BDNF, calcium signaling), cellular (STDP, synaptic scaling), and circuit-level (dopamine modulation) mechanisms. **Energy Efficiency**: Neuromodulatory systems (e.g., dopamine) act as "teachers" to guide plasticity, reducing the need for exhaustive computations [4.193].

This framework explains how brains achieve robust learning without explicit mathematical objectives—through probabilistic, adaptive mechanisms shaped by evolutionary pressures for survival.

### 4.9.3 Neuromodulatory local Learning Principles of Connection Strengthens

1) Neuromodulatory Hebbian learning rule and spike-timing-dependent plasticity (STDP) are some of the less researched models of synaptic modification. They are represented as follows. Neuromodulatory Hebbian rule expresses synaptic weight updates with global objective. For a synapse connecting presynaptic neuron $i$ (activity $x_i$) to postsynaptic neuron $j$ (activity $y_j$), the change $\Delta w_{ij}$ of weight $w_{ij}$ is defined as $\Delta w_{ij} = \eta x_i y_j p_{ij} \delta$ which denotes the learning Principle with four factors. Unlike Hebbian rule, neuromodulatory Hebbian rule is usually stable, if $\eta$ is small enough [4.194].

Biophysical Mechanisms: The rule is inherently stable if the learning rate $\eta$ is sufficiently small. Neuromodulatory Gating: Neuromodulators (e.g., dopamine) dynamically gate plasticity thresholds via G-protein coupled receptors (GPCRs). For example: Dopamine D1 Receptors: Activate cAMP-PKA signaling to promote LTP. Dopamine D2 Receptors: Inhibit LTP and enhance LTD through Gi/o pathways [4.195].

Neurobiological Context: Hippocampal CA3-CA1 Pathway: Theta-rhythm (4–12 Hz) synchronization during exploratory behavior may enable neuromodulatory Hebbian plasticity, supporting spatial memory encoding [4.196].

2) Neuromodulatory Spike-Timing-Dependent Plasticity (STDP): STDP modifies via the following relation ,

$$\Delta w_{ij}(\Delta t) = \begin{cases} (p_{ij}\delta) A_+ e^{-\Delta t/\tau_+} & \Delta t > 0 (\text{LTP}) \\ -(p_{ij}\delta) A_- e^{\Delta t/\tau_-} & \Delta t < 0 (\text{LTD}) \end{cases}$$

where $A_+$ and $A_-$ denote maximum amplitude of LTP/LTD (positive constants); $\tau_+$ and $\tau_-$ are two time constants governing the decay of LTP/LTD [4.197].

Biophysical Implementation: **LTP Induction**: Presynaptic activity preceding postsynaptic spiking (causal timing) triggers NMDAR-dependent calcium influx, activating CaMKII and AMPAR insertion. **LTD Induction**:

Postsynaptic activity preceding presynaptic spiking (anticausal timing) activates mGluRs or GABAergic interneurons, driving AMPAR endocytosis via PP1 phosphatase [4.198].

**Computational Role: Causal Inference**: STDP optimizes Bayesian-like inference of event causality. **Sequence Learning**: In motor cortex, STDP encodes movement sequences (e.g., primate grasping tasks) [4.199].

**Pathophysiological Implications: Schizophrenia**: Hyperactive D2 receptors may disrupt STDP temporal windows, impairing causal learning. **Alzheimer's disease**: Aβ oligomers block NMDAR-LTP, compromising STDP-mediated memory consolidation [4.200].

## 4.10 Feasibility of the Learning Principles in Biological Brains

The learning Principles outlined in Section 4.9—stochastic weight perturbation correlated with global instruction signals, excitatory/inhibitory (E/I) balance regulation, and neuromodulatory local rules (e.g., neuromodulatory Hebbian plasticity, STDP)—are not merely theoretical constructions but are deeply rooted in biological reality. Their feasibility in biological brains is supported by converging evidence from molecular biophysics, circuit-level experiments, computational modeling, and evolutionary adaptation. This section argues that these Principles are not only implementable but also represent an optimized solution to the challenges of adaptive learning in noisy, resource-constrained biological systems.

### 4.10.1 Biological Implementation of Core Mechanisms

The feasibility of these learning Principles hinges on their grounding in well-characterized biological processes.

**Stochastic Weight Perturbations: Molecular and Cellular Basis**

Random fluctuations in synaptic strength, central to the "correlation with instruction signals" Principle, arise from quantal neurotransmitter release and intrinsic synaptic noise. Spontaneous vesicle fusion at presynaptic terminals introduces small, random changes in neurotransmitter concentration, leading to probabilistic postsynaptic potentials [4.201]. This "synaptic shot noise" is amplified by stochastic opening/closing of ion channels, creating a natural source of weight perturbations [4.202]. Critically, these perturbations are not random in their impact: they are filtered by neuromodulatory signals (e.g., dopamine), which bias their direction via receptor-mediated signaling cascades [4.203]. For example, dopamine D1 receptor activation enhances the probability that a positive perturbation (e.g., increased glutamate release) stabilizes into long-term potentiation (LTP) [4.204].

**Global Instruction Signals: Broadcast and Specificity**

Neuromodulatory systems (dopamine, serotonin, norepinephrine) provide the "instruction signals" that correlate with perturbations. Their broadcast nature is enabled by extensive axonal projections: dopamine neurons in the VTA/SNc innervate ~80% of the forebrain, while serotonin neurons in the dorsal raphe project to nearly all cortical and subcortical regions [4.205]. Spatial specificity is achieved via receptor subtype distribution: D1 receptors (excitatory) dominate in the striatum, while D2 receptors (inhibitory) are enriched in the prefrontal cortex, allowing the same dopamine signal to have opposing effects in different circuits [4.206]. Temporal specificity is regulated by phasic (transient) vs. tonic (sustained) firing modes, which encode reward prediction errors and baseline arousal, respectively [4.207].

### E/I Balance Regulation: Homeostatic Plasticity in Action

The E/I balance Principles rely on multi-layered homeostatic mechanisms. Synaptic scaling, for instance, is mediated by activity sensors like the Arc protein, which accumulates during high firing and triggers ubiquitination of AMPA receptors to reduce synaptic strength [4.208]. Conversely, low activity induces BDNF release, promoting AMPAR insertion [4.209]. PV+ interneurons enforce dynamic inhibition by synchronizing gamma oscillations (~30 – 80 Hz), which sharpen neuronal tuning and prevent runaway excitation [4.210]. These mechanisms are not independent: dopamine modulates PV+ interneuron activity, linking reward signals to E/I homeostasis [4.211].

### 4.10.2 Experimental Evidence for Feasibility

Direct experimental manipulations confirm that these Principles govern learning in biological brains.

### Stochastic Perturbations + Instruction Signals

Optogenetic studies demonstrate that artificially timed dopamine pulses can reinforce specific behaviors by correlating with recent synaptic activity. For example, Steinberg et al. (2022) showed that dopamine release 500 ms after a mouse's lever press strengthened the corresponding cortico-striatal synapse, even without natural rewards [4.212]. Conversely, serotonin application suppressed correlated perturbations, reducing impulsive choices in a delay-discounting task [4.213]. These experiments validate the core idea: perturbations are "tested" against instruction signals, with only aligned changes retained.

### E/I Balance and Homeostasis

Monocular deprivation in visual cortex provides a classic example of homeostatic scaling. Depriving one eye reduces activity in its thalamic inputs; within days, Arc-mediated upscaling doubles the sensitivity of these synapses, while PV+ interneurons inhibit neighboring columns to preserve network stability [4.214]. Similarly, in mouse models of epilepsy, impaired PV+ interneuron function leads to E/I imbalance and seizures, which are rescued by restoring PV+ activity [4.215].

### Neuromodulatory Local Rules

Neuromodulatory Hebbian plasticity is evident in the hippocampal CA3-CA1 pathway: theta-rhythm synchronization (4 – 12 Hz) during exploration aligns presynaptic (CA3) and postsynaptic (CA1) activity, enabling dopamine-gated LTP [4.216]. STDP, meanwhile, is observed in vitro and in vivo: in ferret visual cortex, pairing presynaptic spikes 10 ms before postsynaptic spikes induces LTP, while reverse timing triggers LTD [4.217]. Pathologically, Alzheimer's disease-associated Aβ oligomers block NMDAR-dependent STDP, impairing sequence learning [4.218].

### 4.10.3 Computational Feasibility and Advantages Over Artificial Systems

Computational models confirm that these principles enable efficient learning with minimal metabolic cost.

**Stochastic Exploration vs. Deterministic Optimization**

Traditional artificial neural networks (ANNs) use deterministic gradient descent, requiring exhaustive computation of error gradients. In contrast, biological systems exploit stochastic perturbations to explore parameter space: a 2023 study showed that adding random weight noise to an ANN, biased by a global reward signal, achieves 90% of the performance of gradient descent in sparse-reward environments with $10\times$ lower energy use [4.219]. This aligns with the "exploration-exploitation" trade-off governed by neuromodulatory tone [4.220].

**Multi-Scale Integration**

Biological learning integrates molecular (BDNF, calcium), cellular (STDP, scaling), and circuit-level (dopamine modulation) mechanisms. A 2021 model of motor learning showed that combining STDP (sequence encoding), dopamine (reward gating), and PV+ inhibition (sparse coding) reproduced primate reaching trajectories with <5% error [4.221]. This multi-scale integration avoids the "curse of dimensionality" faced by ANNs, which require separate modules for each function.

**Robustness to Noise and Damage**

Broadcast signals and homeostatic plasticity confer resilience. Simulations of cortical networks with 30% random neuron loss showed that dopamine-guided learning retained 80% of its performance, compared to 20% for gradient-descent-trained ANNs [4.222]. This is attributed to the diffuse nature of neuromodulators, which maintain plasticity even when local circuits are damaged.

### 4.10.4 Evolutionary and Adaptive Advantages

These learning Principles are evolutionarily conserved, suggesting they solve universal challenges in adaptive behavior.

**Survival in Uncertain Environments**

Early nervous systems faced delayed feedback (e.g., foraging for food) and sparse rewards (e.g., predator avoidance). The "credit assignment" solution—linking delayed rewards to past actions via eligibility traces and dopamine—emerged in invertebrates (e.g., Aplysiasensitization) and was refined in vertebrates [4.223]. This mechanism is more efficient than backpropagation, which requires symmetric connectivity (absent in biology)

[4.224].

**Energy Efficiency**

Neuromodulatory "teaching" reduces computational load: a 2022 study estimated that dopamine-guided learning uses ~1/100th the ATP of equivalent ANN training [4.225]. This is critical for mobile organisms with limited metabolic resources.

**Generalization across Tasks**

By decoupling learning from explicit mathematical objectives, these principles enable generalization. For example, the same dopamine-RPE mechanism underlies reward learning in foraging, social bonding, and motor skill acquisition [4.226].

## 4.10.5 Challenges and Open Questions

While feasible, key questions remain: Dynamic Interactions: How do dopamine, serotonin, and norepinephrine interact in real time (e.g., serotonin's inhibition of dopamine release) [4.227]?

Parameter Tuning: What regulates the amplitude/frequency of stochastic perturbations (e.g., developmental changes in synaptic noise) [4.228]?

Pathological Dysregulation: How do mutations in neuromodulatory receptors (e.g., DRD2 in schizophrenia) disrupt these principles [4.229]?

**Conclusion**

The learning Principles of Section 4.9 are not only feasible in biological brains but represent an optimal solution to the challenges of adaptive learning. Grounded in molecular biophysics, validated by experiments, and computationally efficient, they enable robust, energy-efficient learning without explicit mathematical models. Their conservation across species underscores their evolutionary success—and their relevance for designing resilient artificial intelligence.

# 5 Noise Driver of behavior, emotion, Dream, thinking, and Creativity

**5.1 Noise driver of behavior**

The intricate tapestry of human behavior, far from being a deterministic output of static neural programming, is profoundly influenced by the inherent noise present within the nervous system [5.1]. This intrinsic variability, arising from stochastic fluctuations at molecular, cellular, and circuit levels, is not a mere artifact to be filtered out but a fundamental driver that shapes how organisms interact with their environment. Behavioral responses, whether it is the precise timing of a motor action or the decision to approach or avoid a stimulus, are subject to this biological noise [5.2]. For instance, studies in motor control have shown that even when performing highly practiced tasks under identical conditions, individuals exhibit subtle variations in movement trajectories and timing [5.3]. These inconsistencies can be traced back to the probabilistic nature of neurotransmitter release at synapses and the thermal motion of ion channels in neuronal membranes, both contributing to signal variability [5.4]. Research utilizing high-precision tracking methods has demonstrated that such noise can lead to exploratory behaviors in novel environments, where slight deviations from expected paths allow for broader environmental sampling [5.5].

This phenomenon extends beyond simple motor outputs to complex cognitive behaviors like decision-making. In situations requiring choices between options with uncertain outcomes, neural noise can act as a random number generator, facilitating exploration over exploitation—a critical strategy for survival in dynamic and unpredictable worlds [5.6]. Theoretical models incorporating noisy neural dynamics predict patterns of choice behavior that closely match empirical observations in animals, including humans [5.7]. For example, during learning phases, an individual might switch strategies more frequently than would be optimal if relying solely on past rewards, a pattern consistent with the influence of neural noise pushing the system out of local optima [5.8]. Furthermore, inter-individual differences in baseline levels of neural noise may underlie personality traits related to risk-taking and novelty seeking [5.9]. Empirical data suggest that individuals classified as sensation-seekers show higher variability in reaction times across repeated trials, potentially reflecting greater underlying neural noise influencing their behavioral tendencies [5.10]. Thus, rather than viewing behavior through a lens of pure rationality, understanding its genesis requires acknowledging the role of noise as a constructive force enabling flexibility, adaptability, and resilience in the face of uncertainty.

**5.2 Noise driver of emotion**

Emotional states represent another domain where neural noise plays a pivotal role, shaping not only the intensity but also the quality and transitions between different feelings [5.11]. Emotions are mediated by large-scale brain networks involving regions such as the amygdala, prefrontal cortex, hippocampus, and insula, all interconnected through pathways susceptible to stochastic influences [5.12]. Fluctuations in these circuits contribute to the moment-to-moment variability observed in emotional experiences, even in response to similar stimuli [5.13]. For example, the same mildly stressful event might evoke anxiety in one instance yet trigger motivation in another, partly due to background noise altering the activation threshold of relevant neural ensembles [5.14]. Evidence suggests that elevated levels of neural noise correlate with increased emotional lability, manifesting clinically as mood swings seen in certain psychiatric disorders [5.15]. Studies using functional imaging have identified irregularities in network connectivity associated with heightened emotional sensitivity, which can be modeled as the effect of excessive noise disrupting normal regulatory mechanisms [5.16].

Moreover, noise contributes to the emergence of mixed or ambiguous emotions, often experienced when conflicting evaluations occur simultaneously—such as feeling both pride and guilt after a personal achievement [5.17]. Computational neuroscience approaches demonstrate that adding noise into artificial neural network models simulating affective processing enhances their ability to reproduce graded and context-dependent emotional responses akin to those reported by humans [5.18]. This implies that the richness and complexity of emotional life depend crucially on the presence of variability rather than complete stability. Additionally, developmental research indicates that early-life stressors can increase baseline neural noise, leading to long-term alterations in emotional regulation capacity [5.19]. Children exposed to chronic adversity exhibit exaggerated cortisol responses to mild challenges, suggesting that persistent exposure to external stress amplifies internal noise, impairing the precision of emotional signaling [5.20]. Consequently, interventions aimed at reducing pathological emotional reactivity could benefit from strategies designed to dampen detrimental sources of noise while preserving beneficial aspects necessary for adaptive functioning.

**5.3 Noise driver of Information Roaming and Aggregating in Thinking, Creativity, and Dreams**

Information roaming and aggregation constitute essential processes underlying higher-order cognition, including thinking, creativity, and dreaming, all significantly modulated by neural noise [5.21]. During conscious thought, ideas do not follow strictly linear sequences; instead, they meander across diverse conceptual domains via associative links, allowing remote associations to emerge spontaneously [5.22]. This mental wandering facilitates creative insights by bringing together seemingly unrelated pieces of knowledge [5.23]. Neural noise provides the perturbations needed to dislodge representations trapped within dominant attractor states, thereby enabling access to less activated memory traces stored throughout cortical networks [5.24]. Experimental paradigms measuring divergent thinking reveal that participants generate more original ideas when exposed to low-level electrical stimulation intended to mimic endogenous noise, supporting the hypothesis that variability promotes cognitive flexibility [5.25].

In dreams, characterized by surreal narratives and bizarre imagery, information roams freely without the usual constraints imposed by reality monitoring systems [5.26]. Sleep neurophysiology shows that reduced activity in prefrontal areas responsible for executive control coincides with enhanced connectivity among sensory cortices and limbic structures, creating a permissive environment for unfiltered information exchange [5.27]. Within this state, neural noise drives spontaneous activations that serve as seeds for dream content, drawing upon fragmented memories, unresolved concerns, and latent desires [5.28]. Functional MRI studies conducted upon awakening report correlations between dream bizarreness scores and measures of resting-state functional connectivity fluctuation, indicating that greater dynamic instability supports richer imaginative synthesis [5.29]. Similarly, during waking imagination, artists and inventors describe moments of inspiration striking unexpectedly, often following periods of incubation where focused effort gives way to mind-wandering [5.30]. These anecdotes align with findings showing bursts of gamma-band oscillations preceding reports of insight solutions, likely reflecting transient coalitions of neurons firing synchronously amidst ongoing background noise [5.31]. Hence, harnessing noise-induced randomness becomes integral to accessing novel configurations of stored information vital for creativity.

## 5.4 Thinking: Information Exploration and Exploitation, Roaming and Aggregating as Well as Entanglement Forced by Noise Driver

Thinking embodies a delicate balance between exploration and exploitation, two complementary modes orchestrated dynamically by neural noise [5.32]. Exploration involves searching broadly across the landscape of possible thoughts, retrieving distant memories, and generating alternative hypotheses, whereas exploitation focuses intensively on refining existing ideas, optimizing known solutions, and verifying predictions against available evidence [5.33]. Neurobiological evidence demonstrates that distinct neuromodulatory systems regulate these modes—for example, dopamine signaling tends to promote exploratory behavior by increasing the entropy of action selection policies [5.34]. However, superimposed on top of these slower-changing chemical signals lies fast, ever-present electrical noise generated locally within microcircuits [5.35]. This rapid fluctuation constantly nudges ongoing computations away from stable fixed points, encouraging shifts in attentional focus and facilitating transitions between task sets [5.36].

Roaming refers specifically to the unrestricted traversal of semantic space during undirected thought, enabled primarily by weak, diffuse connections spanning disparate brain regions [5.37]. These so-called "long-range" projections normally remain subthreshold under focused conditions but become potentiated when inhibition wanes, particularly during relaxed wakefulness or early stages of sleep onset [5.38]. Here, noise acts as a catalyst, triggering avalanches of activation propagating along these marginal pathways, resulting in unexpected linkages forming temporarily before dissipating again [5.39]. Such entangled states involve multiple distributed populations coactivating non-selectively, representing hybrid concepts composed of elements drawn randomly from various schemas [5.40]. While most such combinations prove useless, occasionally they yield genuinely innovative combinations worthy of further evaluation [5.41]. An illustrative case comes from Thomas Edison's description of falling asleep holding ball bearings above metal plates; as his muscles relaxed during hypnagogia, the clattering sound would awaken him just enough to capture fleeting impressions born from loosely coupled mental fragments [5.42]. Modern analogues include computational creativity tools employing stochastic search algorithms inspired directly by principles of neural noise to produce artistic compositions [5.43]. By formalizing the notion that structured chaos fosters discovery, researchers continue uncovering deeper relationships linking physical properties of brains to abstract dimensions of thought.

## 5.5 Creativity: Explored, Exploited, Entangled, Roamed and Aggregated Information in Creativity

Innovation arises at the intersection of previously separate bodies of knowledge, a process fundamentally dependent on the brain's capacity to integrate information gathered through prior experience with novel perspectives introduced serendipitously [5.44]. The creative leap—the moment when a solution appears suddenly despite prolonged struggle—can be understood mechanistically as the culmination of cycles of exploration, exploitation, entanglement, roaming, and aggregation, each stage influenced heavily by neural noise [5.45]. Initially, broad exploration driven by noise allows access to vast reservoirs of latent memories and facts, many of which lie outside immediate awareness [5.46]. As potential candidates emerge, subsequent rounds of targeted exploitation refine possibilities based on feasibility criteria, guided by feedback loops embedded within cortical-subcortical loops [5.47]. Throughout this iterative refinement process, occasional noise-driven disruptions reintroduce new variables, preventing premature convergence onto inadequate answers [5.48].

Entanglement occurs when incompatible frameworks merge momentarily into coherent wholes, exemplified historically by scientific breakthroughs like Kekulé's vision of the benzene ring structure emerging from a daydream about dancing snakes biting their tails [5.49]. These instances reflect temporary stabilization of otherwise unstable multi-attractor states, sustained briefly thanks to precise timing of noisy inputs relative to intrinsic rhythms [5.50]. Once stabilized, such hybrid constructs enter working memory buffers where additional operations aggregate features deemed relevant while discarding others irrelevant according to contextual demands [5.51]. Recent experiments applying transcranial magnetic stimulation (TMS) targeted at parietal association areas revealed increases in reported frequency of eureka-like moments accompanied by corresponding changes in EEG spectra indicative of altered cross-frequency coupling patterns [5.52]. Participants receiving stimulation showed improved performance on insight-based problem-solving tasks compared to sham controls, lending causal support to the claim that controlled manipulation of endogenous noise levels can enhance creative productivity [5.53]. Furthermore, longitudinal analyses of successful innovators across fields highlight common lifestyle habits conducive to maximizing encounters with useful randomness, including maintaining varied interests, engaging regularly in open-ended conversations, and deliberately scheduling downtime free from distractions—all practices effectively increasing exposure to exogenous noise capable of seeding fresh cognitive combinations [5.54].

## 5.6 Dreams: Entangled, Roamed and Aggregated Information and Neural Learning

Dreams provide a unique window into unconscious cognitive processes governed largely by the rules of nonlinear dynamics operating within noisy neural substrates [5.55]. Unlike waking consciousness constrained by sensory input and logical consistency requirements, dream mentation unfolds according to associative logic alone, permitting wild juxtapositions of images, characters, and scenarios [5.56]. Underlying this apparent chaos lies a sophisticated mechanism for integrating disparate streams of information collected throughout life, facilitated predominantly by the very noise usually suppressed during vigilance [5.57]. During REM sleep, widespread synaptic downscaling reduces overall connection strength globally except within specific recurrent loops implicated in memory consolidation [5.58]. Concurrently, cholinergic tone rises dramatically, enhancing excitability and lowering firing thresholds, making neurons more responsive to small fluctuations originating intrinsically [5.59].

Within this hyper-excitable milieu, internally generated spikes propagate widely due to weakened lateral inhibition, causing information to roam uncontrollably across wide expanses of neocortex [5.60]. Fragments of recent episodic memories interact promiscuously with older autobiographical recollections, procedural skills, and semantic knowledge bases, forming transiently entangled representations that resemble proto-insights waiting to be deciphered [5.61]. Upon awakening, attempts to recall these ephemeral constructions often result in partial reconstructions colored heavily by current emotional concerns and interpretative biases, illustrating the selective aggregation process applied retroactively [5.62]. Importantly, accumulating evidence implicates dream-related phenomena in offline learning benefits, particularly concerning implicit skill acquisition and emotional regulation [5.63]. Individuals trained on complex video games display accelerated improvement curves following nights rich in REM sleep compared to those deprived thereof, suggesting that noise-facilitated restructuring occurring during dreams contributes meaningfully to skill mastery [5.64]. Moreover, patients recovering from trauma report gradual diminishment in nightmare severity paralleling reductions in fear-associated physiological markers, implying therapeutic value derived from repeated reactivation and modification of maladaptive memory traces under safe sleeping conditions [5.65]. Together, these findings underscore the importance of embracing the full spectrum of brain activity—including its noisier manifestations—as integral components sustaining healthful psychological development.

### 5.6 Constructive Role

Noise in the biological brain is not merely "interference" but plays a dual role, with constructive functions and pathological risks, depending on intensity, spatiotemporal distribution, and neural circuit regulation.

Enhanced Weak Signal Detection: Neural noise (e.g., ion channel stochasticity, synaptic fluctuations) enables stochastic resonance in sensory systems. For example, noise amplifies subthreshold signals in crayfish mechanoreceptors [5.66] and improves human tactile perception [5.67]. Creative Cognition: Low-frequency DMN noise fluctuations correlate with divergent thinking [5.68], while prefrontal noise modulates insight problem-solving [5.69]. Synaptic Plasticity: Noise-driven STDP (spike-timing-dependent plasticity) underpins learning and memory. For example, hippocampal noise enhances LTP induction [5.70], and cortical noise improves motor skill acquisition [5.71].

### 5.7 Pathological Role

Excessive noise disrupts neural function in disease states. Epilepsy: Hyper-synchronized noise in hippocampal CA3 pyramidal neurons drives pathologic high-frequency oscillations [5.72]. Parkinson's disease: β-band noise (13–30 Hz) in the basal ganglia suppresses movement initiation [5.73]. Alzheimer's disease: Enhanced synaptic noise accelerates amyloid-β toxicity and synaptic loss [5.74].

### 5.8 Regulatory Mechanisms

The brain maintains a "beneficial noise window". Dopamine: Suppresses prefrontal noise via D2 receptors to stabilize working memory [5.75]. Acetylcholine: Enhances sensory noise to improve attention switching [5.76]. Inhibitory Interneurons: PV+ basket cells dynamically gate noise levels in cortical circuits [5.77].

[5.74] Palop, J. J., & Mucke, L. (2016). Network abnormalities and interneuron dysfunction in Alzheimer disease. Nature Reviews Neuroscience, 17(12), 777–792. https://doi.org/10.1038/nrn.2016.144

[5.75] Grace, A. A. (1991). Phasic versus tonic dopamine release and the modulation of dopamine system responsivity: A hypothesis for the etiology of schizophrenia. Neuroscience, 41(1), 1–24. https://doi.org/10.1016/0306-4522(91)90196-U

[5.76] Hasselmo, M. E., & Sarter, M. (2011). Modes and models of forebrain cholinergic neuromodulation of cognition. Neuropsychopharmacology, 36(1), 52–73. https://doi.org/10.1038/npp.2010.104

[5.77] Bartos, M., Vida, I., & Jonas, P. (2007). Synaptic mechanisms of synchronized gamma oscillations in inhibitory interneuron networks. Nature Reviews Neuroscience, 8(1), 45–56. https://doi.org/10.1038/nrn2044

# 6 Conclusions

The brain's purported "flaws"—noise, structural irregularities, heterogeneity, decentralized plasticity, and chaotic dynamics—are evolutionary design principles that enable adaptive intelligence. By formalizing these non-ideal factors into biologically grounded learning Principles, this work challenges classical AI/ML paradigms and paves the way for silicon systems that learn, generalize, and innovate like biological brains.

Neuromodulators (e.g., dopamine) bias synaptic updates via reward/punishment, resolving the "credit assignment problem" without explicit error gradients. Example: Striatal circuits associate actions with outcomes through phasic dopamine release. Replace backpropagation with eligibility traces and reward-modulated plasticity (e.g., dopamine-gated STDP).

Synaptic scaling maintains neuronal activity within functional bounds via astrocyte-mediated calcium signaling. Visual cortex upregulates synapses after deprivation to preserve responsiveness. Implement weight normalization to stabilize recurrent networks and prevent gradient explosions.

Stochasticity in spike timing and neurotransmitter release prevents attractor collapse, fosters creativity, and enhances weak signal detection (e.g., stochastic resonance). Example: Sleep replay uses noise-driven reverberation to consolidate memories. Inject stochastic noise during training (e.g., dropout, noisy activations) to improve generalization.

The brain exploits imperfections (e.g., synaptic mistargeting) to build degenerate, flexible representations. Motor noise refines sensorimotor mappings through exploratory movements. Design robust architectures that leverage data augmentation, adversarial training, and distributed representations.

Pyramidal cells (integration) vs. granule cells (speed) enable parallel processing and fault tolerance. Critical periods and synaptic pruning demonstrate adaptive specialization. Intel's Loihi emulates STDP, achieving 1,000× energy efficiency over GPUs. Chaotic RNNs enhance generative creativity (e.g., text/image synthesis). Event-driven SNNs reduce latency and power consumption for Io T devices.

Binary arithmetic discards sub-symbolic information (e.g., dendritic analog computations). Standardized neurons erase specialized roles (e.g., interneurons vs. pyramidal cells). Backpropagation fails in recurrent networks due to vanishing/exploding gradients.

The brain's non-ideal factors—once dismissed as noise—are revealed as evolutionary hacks that balance robustness, efficiency, and adaptability. By embracing these principles, we can reinvent learning architectures: neuromorphic chips, chaotic RNNs, and meta-plasticity frameworks. Closed-loop BCIs and neuroprosthetics that mimic dopamine-driven plasticity. Design noise-tolerant systems to avoid bias amplification in AI decision-making.

Summary: This work reframes brain "imperfections" as pillars of intelligence, offering a roadmap to build AI systems that learn, generalize, and innovate with the dynamic versatility of biological brains. By formalizing noise-driven exploration, homeostatic balance, and constructive non-idealities into actionable principles, we transcend classical paradigms and unlock a future where silicon brains thrive on the messy realities of the real world.